%% file: tracing_lya.tex
\documentclass[useAMS,usenatbib]{mn2e}

\usepackage{times}  
\usepackage{listings}
\usepackage{graphics}
\usepackage{subfigure}
\usepackage{graphicx}
\usepackage{graphicx,xspace}
\usepackage{amssymb}
\usepackage{amsmath}
\usepackage{aas_macros}
\usepackage{lscape}
\usepackage{rotating}
\usepackage{placeins}
\usepackage{natbib}
\usepackage{color}

\newcommand{\lya}{\mbox{Ly$\alpha$}\xspace}
\newcommand{\ramses}{{\sc{ramses}}\xspace}
\newcommand{\ramsesrt}{{\sc{ramses-rt}}\xspace}
\newcommand{\music}{{\sc{music}}\xspace}

\newcommand{\cloudy}{{\sc{cloudy}}\xspace}

\newcommand{\sphinx}{{\sc{sphinx}}\xspace}
\newcommand{\bpass}{{B\sc{pass}}\xspace}
\newcommand{\rascas}{{\sc{rascas}}\xspace}

\title[Tracing the CGM with \lya emission]
{Tracing the simulated high-redshift circum-galactic medium with Lyman $\alpha$ emission}

\author[P. D. Mitchell et al.]{
\newauthor Peter D. Mitchell\thanks{\rm E-mail: mitchell@strw.leidenuniv.nl}$^{1}$,
J\'{e}r\'{e}my Blaizot$^{2}$,
Corentin Cadiou$^{3}$,
Yohan Dubois$^{4}$, 
\newauthor Thibault Garel$^{5,2}$, and
Joakim Rosdahl$^{2}$
\\
$^{1}$Leiden Observatory, Leiden University, P.O. Box 9513, 2300 RA Leiden, the Netherlands\\
$^{2}$Univ Lyon, Univ Lyon1, Ens de Lyon, CNRS, Centre de Recherche Astrophysique de Lyon UMR5574, F-69230, Saint-Genis-Laval, France\\
$^{3}$Department of Physics and Astronomy, University College London, Gower Street, London WC1E 6BT, UK\\
$^{4}$Institut d’Astrophysique de Paris, Sorbonne Universit\'{e}, CNRS, UMR 7095, 98 bis bd Arago, 75014 Paris, France\\
$^{5}$Observatoire de Gen\`{e}ve, Universit\'e de Gen\`{e}ve, 51 chemin de P\'egase, 1290 Versoix, Switzerland\\
}

\begin{document}
\date{\today}
\pagerange{\pageref{firstpage}--\pageref{lastpage}} \pubyear{2020}
\maketitle
\label{firstpage}

\begin{abstract}
With the Multi Unit Spectroscopic Explorer (MUSE), it is now possible to detect spatially extended Lyman $\alpha$ (\lya) emission 
from individual faint ($M_{\mathrm{UV}} \sim -18$) galaxies at redshifts, $3 < z < 6$, tracing gas out to 
circum-galactic scales comparable to the dark matter halo virial radius.
To explore the implications of such observations, we present a cosmological radiation hydrodynamics simulation 
of a single galaxy, chosen to be typical of the \lya-emitting galaxies detected by MUSE in deep fields.
We use this simulation to study the origin and dynamics of the high-redshift circum-galactic medium (CGM).
We find that the majority of the mass in the diffuse CGM is comprised of material infalling for the first time towards the halo center, 
but with the inner CGM also containing a comparable amount of mass that has moved past first-pericentric passage,
and is in the process of settling into a rotationally supported configuration. 
Making the connection to \lya emission, we find that the {\it observed} extended surface brightness profile 
is due to a combination of three components: scattering of galactic \lya emission in the CGM, in-situ 
emission of CGM gas (mostly infalling), and \lya emission from small satellite galaxies. The weight of 
these contributions vary with distance from the galaxy such that (1) scattering dominates the inner regions 
($r < 7 \, \mathrm{kpc}$), at surface brightness larger than a few $10^{-19}$ cgs, (2) all components contribute 
equally around $r \sim 10 \, \mathrm{kpc}$ (or SB$\sim 10^{-19}$), and (3) the contribution of small 
satellite galaxies takes over at large distances (or SB$\sim 10^{-20}$).
Compared to stacked MUSE observations, we show that we can reproduce the observed \lya 
surface brightness profile closely, to within at least $0.2 \, \mathrm{dex}$, spanning over
 two orders of magnitude in surface brightness, and out to beyond the halo virial radius.
We find that the CGM produces on average slightly blue-shifted (and single peaked)
\lya spectra, reflecting the net inflow of dense circum-galactic gas. 
Our simulation fails to reproduce the characteristic observed \lya spectral morphology that is 
red-shifted with respect to the systemic velocity, with the implication that the
simulation is missing an important component of neutral outflowing gas.
\end{abstract}

\begin{keywords}
galaxies: formation -- galaxies: evolution
\end{keywords}

\vspace*{-0.5cm}  

\section{Introduction}


Spatially extended ``haloes'' of \lya emission are observed around
galaxies across a range of scales, ranging from 
massive radio galaxies and quasars \cite[e.g.][]{Heckman91,Reuland03,Cantalupo14}, 
to UV bright Lyman Break Galaxies \cite[LBGs][]{Steidel11},
down to comparatively UV-faint \lya-emitters (LAEs)
\cite[e.g.][]{Rauch08,Matsuda12,Hayes13,Wisotzki16}.
These observations demonstrate that there must be substantial amounts
of dense hydrogen gas in the circum-galactic
medium (CGM) around high-redshift galaxies. Comparisons to the incidence
rates of strong hydrogen absorbers (specifically Lyman limit systems) 
seen in quasar spectra support the idea that this extended emission is tracing
similar gas \cite[][]{Wisotzki18}, and direct evidence for connection between faint
high-redshift LAEs and strong hydrogen absorbers is now starting to appear 
\cite[][]{Lofthouse19,Mackenzie19,Diaz20,Muzahid20}.

The detection of spatially extended \lya emission around high-redshift
galaxies poses a number of questions. First, what component(s) of the
CGM is being traced? Cosmological simulations predict for example the presence
of inflowing filamentary streams of dense gas \cite[e.g.][]{Keres05,Dekel06}, and 
in addition dense gas could be either entrained or formed in situ within galactic-scale
outflows \cite[e.g.][]{Wang95,Heckman00,Scannapieco17}. 
Second, a related but distinct question is
what is the production mechanism for the extended \lya emission?
An obvious mechanism is that \lya is emitted within ionized HII
regions inside the interstellar medium (ISM), which then resonantly
scatter from neutral hydrogen in the CGM \cite[e.g.][]{Laursen07,Steidel10,Zheng11}. Alternatively,
the observed photons could be emitted within the HII regions of satellite
galaxies, or in situ within the CGM due to photo-ionization by
escaping ionizing photons (fluorescence) originating from the host galaxy or quasar
\cite[e.g.][]{Haiman01,MasRibas16,Gallego18}, from
satellites \cite[e.g.][]{MasRibas16,MasRibas17}, or from the wider ultraviolet background \cite[UVB, e.g.][]{Furlanetto05}, or
because of collisional excitations and recombinations that dissipate
the energy gained by compressive heating, which can be triggered
by cosmological gas infall \cite[dubbed ``cooling radiation'', e.g.][]{Fardal01,Dijkstra06,FaucherGiguere10,Rosdahl12},
or by supernova shock waves that break out of the ISM \cite[][]{Mori04}.

\cite{Steidel10} propose that spatially extended cool gas within galactic 
outflows can explain simultaneously the kinematics of metal absorption
lines detected ``down the barrel'' in the spectra of high-redshift LBGs, the \lya
spectra of LBGs (which are red-shifted from systemic), the equivalent width
of absorbing cool gas as a function of transverse distance (detected via
background sources), and \cite{Steidel11} argue that this picture can
also explain the spatially extended \lya emission from the CGM 
\cite[see also][]{Dijkstra12}.
The question of extended \lya emission aside, simple outflowing neutral gas
shell models are generally used to explain the spectral morphology of observed
\lya spectra \cite[e.g.][]{Ahn04,Verhamme06}. In this scenario, \lya photons escaping a central source
backscatter from side to side of the inner surface of a circum-galactic outflowing shell, until
they shift in frequency sufficiently to be out of resonance with
the blue-shifted side of the shell moving towards the observer.
Outflowing shell models can explain the asymmetry and often
complex spectral morphology of observed \lya lines, the commonly
observed single peaked \lya line that is red-shifted from the systemic
velocity, and why there is often less (or no) emission observed at (and blueward of)
the systemic velocity \cite[e.g.][]{Verhamme08,Wofford13,Hashimoto15,Gronke17b}.
Similar models with a more continuous expanding medium can also explain
observed surface brightness profiles \cite[e.g.][]{Song20}.

These observations do not rule out the other scenarios however.
The spectral morphology of observed \lya lines could be imprinted
on small scales within the ISM, rather than on CGM scales. 
Similarly, the spatial location of strongly outflowing gas seen down-the-barrel
in galaxy spectra is highly uncertain, and could be tracing be sub-ISM
scales of order $\sim 100 \, \mathrm{pc}$ \cite[][]{Chisholm16},
and so not connected to the absorbing gas seen in transverse sight-lines
(where the kinematics relative to the galaxy are poorly constrained).
As such, the door remains open for other CGM components to provide
the main explanation for the observed spatial distribution of \lya haloes,
without the CGM necessarily being responsible for the observed \lya
spectral morphology.

A natural way to explore these questions is with cosmological simulations,
which in principle can simultaneously capture the complex mix of
circum-galactic gas flows that surround galaxies, sourced by the
filamentary infall of matter from larger scales. It must
be acknowledged that the finite resolution of simulations 
\cite[particularly in the CGM, see][]{Hummels19,Peeples19,Suresh19,vandevoort19} limits their ability to accurately represent
multi-phase gaseous media; the resolution required to resolve
the formation of dense clumps in the CGM or in winds has been
estimated to be as small as $0.1 \, \mathrm{pc}$ \cite[e.g.][]{Fujita09,Mccourt18}.
Nonetheless, many studies
of \lya emission in connection to high-redshift galaxies
have been performed with cosmological simulations 
\cite[e.g.][]{Cantalupo05,Tasitsiomi06,Laursen07,Zheng10,Barnes11}, 
with the simulations progressively increasing in both complexity and resolution.
Simulations have been used to explore the effect of scattering 
of centrally emitted \lya photons from the CGM to create extended emission
\cite[e.g.][]{Barnes11,Zheng11,Lake15}, to study in-situ gravitational ``cooling'' radiation 
\cite[e.g.][]{FaucherGiguere10,Goerdt10,Rosdahl12},
as well as fluorescent emission powered by the ultraviolet background (UVB), and by proximate
quasars \cite[e.g.][]{Cantalupo05,Kollmeier10}.

\cite{Verhamme12} use non-cosmological high-resolution simulations of 
idealised galaxies and emphasise the importance of allowing a cold ISM
phase to form for studying \lya escape from the ISM.
Studies of \lya transfer in cosmological simulations of high-redshift galaxies
with the resolution to start addressing this challenge are now starting to appear. 
\cite{Behrens19} study
\lya escape from a $z \sim 8$ galaxy with maximum spatial resolution
of $25 \, \mathrm{pc}$ and gas mass resolution of $2 \times 10^4 \, \mathrm{M_\odot}$,
employing full \lya radiative transfer within the ISM and CGM, but using
simplified modelling to account for photo-ionization by local sources.
They find generally low \lya escape fractions as a result of strong local dust attenuation
around young stellar clusters.
Notably, \cite{Smith19} analyse a galaxy at $z=5-7$ from the FIRE-2 suite of
cosmological zoom-in simulations \cite[][]{Hopkins18}, with
a gas mass resolution of $7 \times 10^3 \, \mathrm{M_\odot}$.
They account for photo-ionization from local
sources by post-processing their simulation with ionizing UV photons,
and also perform full \lya radiative transfer through both the ISM and CGM.
They study the angular and temporal variations of escaping \lya emission,
finding that temporal variations are primarily driven simply by the
star formation history of the primary galaxy. They also explore in post-processing
the possible impact of \lya radiation pressure in and around the simulated
galaxy, finding that it should be dynamically important.

Here, we present results from a full cosmological radiation hydrodynamics (RHD) simulation 
of a high-redshift galaxy for $z \geq 3$, taken from the parent sample of
(non-RHD) simulations presented in \cite{Mitchell18}. The simulation self-consistently 
includes the full range of expected \lya powering mechanisms, including
photo-ionization and photo-heating from local sources and the UVB.
The simulation achieves a maximum spatial resolution of $14 \, \mathrm{pc}$
within the ISM at $z=3$,
with a characteristic gas cell (and star particle) mass of $\sim 10^{3} \, \mathrm{M_\odot}$.
This simulation represents a powerful tool to study the radiative
transfer problem for \lya photons in both the ISM and the CGM simultaneously.
Our simulated galaxy (with a halo mass of $M_{200}= 10^{11.1} \, \mathrm{M_\odot}$ at $z=3$)
is deliberately chosen to be representative of the UV-faint LAEs detected
by the Multi Unit Spectroscopic Explorer (MUSE) in deep MUSE fields ($\geq 10$ hour exposure) 
for $3<z<6$ \cite[e.g.][]{Wisotzki16,Hashimoto17,Inami17,Leclercq17}. The substantial
increase in sensitivity afforded by MUSE is significant in this sense,
as this greatly facilitates the computational challenge of producing observable
galaxies in cosmological simulations with high resolution and full radiation 
hydrodynamics.

Building on our previous work presented in \cite{Mitchell18}, we set out
in this study to first understand what shapes the dense phases
of hydrogen in the high-redshift CGM. We then use this information
to interpret predictions from our simulation for spatially extended
\lya emission, and for the spectral morphology of the
\lya line. We address the questions of both how \lya traces the CGM,
and of the origin of the extended emission (in terms of in-situ
versus scattered emission that originates from the host galaxy or from satellites),
albeit we do not undertake here the non-trivial task of
separating the \lya emitted in situ from the CGM between fluorescence
and non-radiative heating sources.
Complementing our study of the CGM, a forthcoming paper (Blaizot et al., in preparation) will
extend the sample of radiation hydrodynamics simulations, and 
will focus on the observability of \lya emission from
the host galaxy, the angular and temporal dependence of the
\lya escape fraction, equivalent width, and spectral morphology,
and the relative role of collisional excitations and recombinations to
the escaping \lya signal.

The layout of this article is as follows: we describe our simulation setup,
subgrid models, and post-processing strategy in Section~\ref{methods},
we present our results in Section~\ref{flows_sec}, we discuss various caveats and 
implications of these results in Section~\ref{discussion_section}, and summarise
in Section~\ref{summary_sec}.

\section{Methods}
\label{methods}

\subsection{Simulation setup}
\label{simul_description}

The results presented in this paper are taken from a single cosmological zoom-in simulation of a high-redshift
galaxy, simulated down to $z=3$ using the \ramsesrt code \cite[][]{Rosdahl13,Rosdahl15}, which is a radiation hydrodynamics extension
of the \ramses code \cite[][]{Teyssier02}. \ramses is an Eulerian code for hydrodynamics that employs adaptive
mesh refinement. The zoom-in simulation presented uses the same subgrid physics models and simulation parameters
as the fiducial \sphinx simulations presented in \cite{Rosdahl18}, and effectively extends the project
from $z=6$ (the final redshift currently achieved in \sphinx) to $z=3$. The initial conditions for the simulation
are taken from the parent sample described in \cite{Mitchell18}, with this simulation representing a resimulation
of the most massive halo from that study, with a halo mass of $M_{200} = 10^{11.1} \, \mathrm{M_\odot}$ and
virial radius of $R_{\mathrm{vir}} = 37 \, \mathrm{kpc}$ at $z=3$.

\defcitealias{Planck13}{Planck Collaboration, 2014}  
We assume an underlying $\Lambda$ Cold Dark Matter cosmological model, with $\Omega_{\mathrm{M}} = 0.3175$, $\Omega_{\mathrm{\Lambda}} = 0.6825$,
$\Omega_{\mathrm{B}} = 0.049$, $H_{0} = 67.11 \, \mathrm{km s^{-1} Mpc^{-1}}$, $n_{\mathrm{s}} = 0.962$, and $\sigma_8 = 0.83$
\citepalias{Planck13}. We use the \music code to generate initial conditions \cite[][]{Hahn11}. The zoom-in simulation
is performed within a low-resolution box of volume $(20 \, \mathrm{cMpc}/h)^3$, with the high-resolution
region selected as a sphere of $150 \, \mathrm{pkpc}$ radius around the target halo at $z=3$.
We assume hydrogen and helium mass fractions of $X=0.76$ and $Y=0.24$ respectively, and we initialise cells
with a starting metallicity of $3.2 \times 10^{-4} \, Z_{\odot}$, assuming the Solar metal mass fraction $Z_{\odot}=0.02$.
This higher-than-pristine metallicity value is chosen to account for the enrichment from stars forming
in unresolved haloes with masses below the atomic cooling limit. 

For the hydrodynamics, we solve the Euler equations using a second order Gudunov scheme, using the Harten-Lax-van Leer-Contact (HLLC)
Riemann solver with a MinMod slope limiter to construct fluid variables at cell interfaces. We employ a Courant factor
of $0.8$ to regulate the simulation timestep, and we assume an adiabatic index of $\gamma=5/3$ to close the
relationship between gas pressure and internal energy, appropriate for an ideal monoatomic gas. Gravitational dynamics are solved using a particle-mesh solver
and cloud-in-cell interpolation \cite[][]{Guillet11}. Advection of radiation between cells is solved
using a first-order moment method, using the M1 closure method \cite[][]{Levermore84} and the Global-Lax-Friedrich
flux function to construct the radiation field at cell interfaces.

\subsubsection{Resolution and refinement strategy}

The simulation uses a dark matter particle mass of $m_{\mathrm{DM}} = 10^{4} \, \mathrm{M_\odot}$ within the zoom-in region.
The base refinement level of the mesh in the zoom-in region is $12$ (proper cell width of $1.8 \, \mathrm{pkpc}$ at $z=3$),
and it is allowed to refine up to a maximum refinement level of $19$ by $z=3$, corresponding to a minimum proper cell
size of $14 \, \mathrm{pkpc}$ at $z=3$. At $z=3$ there are $1.2 \times 10^7$ leaf cells within $R_{\mathrm{vir}}$,
the median cell width in the diffuse CGM ($0.2 < r / R_{\mathrm{vir}} < 1$, excluding satellite galaxies) is 
$227 \, \mathrm{pc}$ (weighted by mass) and $454 \, \mathrm{pc}$ (weighted by volume), and the median cell
size in the ISM ($r < 0.2 R_{\mathrm{vir}}$) is $14 \, \mathrm{pc}$ (weighted by mass) and $113 \, \mathrm{pc}$
(weighted by volume).

Cells are refined either if: \textit{i)} $M_{\mathrm{DM,cell}} + \frac{\Omega_{\mathrm{M}}}{\Omega_{\mathrm{B}}} M_{\mathrm{baryon,cell}} > 8 \, m_{\mathrm{DM}}$,
where $M_{\mathrm{DM,cell}}$ and $M_{\mathrm{baryon,cell}}$ are respectively the total dark matter and baryonic mass enclosed
within a cell, \textit{ii)} the cell width is larger than one quarter of the Jeans length $\lambda_{\mathrm{J}} \equiv \sqrt{\frac{\pi c_{\mathrm{s}}^2}{G \rho_{\mathrm{gas}}}}$,
where $c_{\mathrm{s}}$ is the gas sound speed, $\rho_{\mathrm{gas}}$ is the gas mass density, and $G$ is the gravitational constant.
We do not impose an equation of state to artificially pressurize the ISM if the Jeans length is unresolved, which
can occur when the Jeans length is smaller than the minimum allowed cell size.

\subsubsection{Radiation and thermochemistry}

To represent the locally produced ultraviolet (UV) radiation field, we employ three discrete radiation bins, with frequencies bracketed by the 
ionization energies of hydrogen and helium, including both the first and second ionization state. Full non-equilibrium
interactions between ionizing UV photons, hydrogen and helium are modelled on the fly as described in \cite{Rosdahl13,Rosdahl15}, including the effects
of photo-ionization, heating and momentum transfer. The gas temperature and ionization states of hydrogen and helium are
all tracked and updated accordingly. 
Radiation advection is subcycled relative to the hydrodynamical timestep,
and we use a reduced speed of light approximation with an effective advection velocity of $1/80$ times the true speed of light.
As described in \cite{Rosdahl18}, we use \bpass stellar evolution models \cite[][]{Elridge08} to set the input spectral
energy distribution of star particles. We use \bpass version 2.0 \cite[][]{Elridge16,Stanway16} which assumes
a $100 \, \%$ binary fraction. Since our zoom-in simulation targets only a single galaxy, we include the effects of a homogeneous 
evolving ultraviolet background (UVB) radiation field, in addition to the radiation provided by local stellar sources. We adopt the 
model for the UVB from \cite{FaucherGiguere09}, but apply a self-shielding correction that exponentially damps the UVB intensity 
for gas cells with $n_{\mathrm{H}} > 10^{-2} \, \mathrm{cm^{-3}}$ \cite[][]{FaucherGiguere10,Rosdahl12}. 

In addition to the full non-equilibrium thermochemistry solved for hydrogen and helium, metal line cooling is modelled
for gas with temperature $T>10^4 \, \mathrm{K}$ using \cloudy \cite[][]{Ferland98}, assuming ionization equilibrium
with a \cite{Haardt96} UVB (and so is not modelled self-consistently with the non-equilibrium hydrogen and helium thermochemistry). 
For $T < 10^4 \, \mathrm{K}$ we apply fine structure cooling rates from \cite{rosen95}.

\subsubsection{Star formation}

We use a model for star formation inspired by \cite{Federrath12}, as described in \cite{Kimm17}, \cite{Trebitsch17}, and
\cite{Rosdahl18}. The model accounts for the stabilising effects of both thermal pressure and turbulent motions against
gravitational collapse. The precise criteria for stars to form in a cell are that: \textit{(i)}, the local hydrogen number
density $n_{\mathrm{H}} > 10 \, \mathrm{cm^{-3}}$ (and that the local overdensity is greater than $200$ times the cosmic mean), 
\textit{(ii)}, the gas is locally convergent and the host cell represents a local density maxima compared to the six nearest cells,
\textit{(iii)}, the cell width $\Delta x$ is smaller than the turbulent Jeans length given by

\begin{equation}
\lambda_{\mathrm{J,turb}} = \frac{\pi \sigma_{\mathrm{gas}}^2 + \sqrt{36 \pi c_{\mathrm{s}}^2 G \Delta x^2 \rho + \pi^2 \sigma_{\mathrm{gas}}^4 }}{6 G \rho \Delta x},
\end{equation}

\noindent where $\sigma_{\mathrm{gas}}$ is the local gas velocity dispersion,
computed using the velocity gradients to cells that share vertices with the host cell. As described in \cite{Rosdahl18},
the computation of $\sigma_{\mathrm{gas}}$ is modified with respect to the model described in \cite{Kimm17} by subtracting
rotational and symmetric divergent components of the velocity field, such that the turbulent support against a converging flow
is only provided by the relevant anisotropic motions.

For cells that meet these criteria, star particles are allowed to form stochastically with (on average) a rate given by
$\dot{\rho}_\star = \epsilon_\star \rho_{\mathrm{gas}} / t_{\mathrm{ff}}$, where 
$t_{\mathrm{ff}} = [3 \pi / (32 G \rho_{\mathrm{gas}})]^{1/2}$ is the local freefall time, and $\epsilon_\star$ is
a variable efficiency that depends on the local turbulent properties of the gas and on the sound speed 
\cite[see][for details]{Kimm17,Trebitsch17}. This efficiency can often reach or exceed values of unity, leading
to a bursty scenario for star formation.

\subsubsection{Stellar evolution and supernova feedback}
\label{sec_sf_fb}

We model the effects of supernova explosions using the mechanical feedback scheme described in \cite{Kimm14} 
and \cite{Kimm15}; specifically we use the mechanical feedback scheme with multiple explosions
from the list of variant models considered in \cite{Kimm15}. For each star particle with age $<50 \, \mathrm{Myr}$ and $> 3 \, \mathrm{Myr}$, a stochastic sampling
is performed over the supernova delay time distribution, with individual supernova explosions
of $10^{51} \, \mathrm{erg}$ that inject energy/momentum into the host cell and neighbouring cells around the particle.
We assume a \cite{Kroupa02} stellar initial mass function, with $20 \%$ of the star particle mass 
returned to the neighbouring cells across all supernova events, and in addition metals are injected with a metal
yield $y=0.075$. As discussed in \cite{Rosdahl18} and \cite{Mitchell18}, we break consistency with the Kroupa stellar initial mass function (IMF) to decide how
many supernova explosions occur for each star particle, increasing the number by a factor four such that there
are four supernova explosions (each of $10^{51} \, \mathrm{erg}$) per $100$ Solar masses of stars formed. This choice was
made by \cite{Rosdahl18} to calibrate the \sphinx simulations against observational constraints for $z \geq 6$
and was made by \cite{Mitchell18} to improve agreement with constraints for $z \geq 3$.

Depending on the cell metallicities and densities, the mechanical scheme determines for each cell whether the simulation is able
to resolve the adiabatic phase (in which case the injected energy is $10^{51} \, \mathrm{erg}$ and the injected
momentum scales $\propto \sqrt{m}$, where $m$ is the mass of the neighbouring cells that is considered to be 
entrained into the outflow), and otherwise injects energy and the maximum momentum achieved in the
momentum-conserving snowplow phase, with a maximum injected momentum of $3 \times 10^5 \, \mathrm{M_\odot kms^{-1}}$
for gas with $n_{\mathrm{H}} = 1 \, \mathrm{cm^{-3}}$ and $Z=Z_\odot$ \cite[][]{Blondin98,Thornton98}.
Following \cite{Kimm17}, we then boost the maximum momentum to $4.2 \times 10^5 \, \mathrm{M_\odot kms^{-1}}$
if the local Stromgren radius is unresolved, accounting for unresolved photo-ionizing heating from the star particle, 
which can reduce the densities around the star and so reduce the radiative losses \cite[][]{Geen15}.

\subsection{Halo finder and satellite identification}

We identify both central and satellite dark matter subhaloes using the AdaptaHOP algorithm \cite[][]{Aubert04}, and construct
merger trees following \cite{Tweed09}. We store simulation outputs with a temporal cadence of $10 \, \mathrm{Myr}$.
Halo centres are defined based on the dark matter density
maxima. All halo masses ($M_{200}$) and virial radii ($R_{\mathrm{vir}}$) quoted are measured by computing the spherical radius within
which the mean enclosed density is $200$ times the critical density of the Universe.

For each satellite identified by the halo finder, we assign particles and gas cells to that satellite if
they fall within the tidal radius, which we approximate following \cite{Binney08} as

\begin{equation}
r_{\mathrm{t}} = R_0 \left(\frac{m}{M(R_0) \left(3 - \frac{\mathrm{d}\ln{M}}{\mathrm{d}\ln(R)}|_{R=R_0}\right)}\right)^{1/3},
\end{equation}

\noindent where $R_0$ is the distance from the satellite centre to the host
centre, $M(R_0)$ is the mass of the host enclosed within $R_0$ and $m$ is the mass of
the satellite (the mass within $r_{\mathrm{t}}$). This relation is appropriate
for a satellite on a circular orbit within a spherically symmetric host potential.

\subsection{Tracer particles and classifying circum-galactic gas}
\label{tracer_methods}

The results presented in this paper make extensive use of tracer particles to provide Lagrangian information
about the trajectories of gas in our simulation. We use the Monte Carlo implementation of tracer particles
introduced by \cite{Genel13}, and adapted for use in \ramses simulations by \cite{Cadiou19}. \cite{Cadiou19}
show that these tracers accurately reproduce the underlying gas density field, and self-consistently follow
our subgrid models for star formation and feedback. We use a relatively high tracer sampling, with ten tracers 
initialised per gas cell in the zoom-in region of the initial conditions. 

One application of tracer particles is to classify the nature of the gas contained within a given cell in the
simulation. We can then use these components to (for example) decompose \lya emission from the CGM into
different components, thereby gaining insight into the CGM-\lya connection.
Unless otherwise stated, we use the following definitions to define different (mutually exclusive) 
gaseous components within the virial radius of the primary host halo:

\begin{itemize}
\item \textit{ISM}: any gas with radius $r < 0.2 \, R_{\mathrm{vir}}$. 
In practice this also includes material that could be argued to belong to an ISM-CGM interface, or to the inner CGM, but this is qualitatively unimportant for our results.
\item \textit{Satellite gas}: any gas that is within the tidal radius of a satellite galaxy.
\item \textit{Stripped CGM}: circum-galactic gas that was within the tidal radius of a satellite (for at least $30 \, \mathrm{Myr}$) 
as it entered the virial radius of the host halo, but is subsequently located beyond the tidal radius.
This could be related to ram pressure stripping, gravitational tidal forces, or stellar feedback from the satellite; we do not attempt to distinguish between these mechanisms.
\item \textit{First-infall CGM}: circum-galactic gas that is radially infalling from the inter-galactic medium (IGM) for the first time.
We account for occasional fluctuations in the radial velocity of tracers by requiring that a tracer be radially outflowing 
consecutively for $30 \, \mathrm{Myr}$ after the current snapshot in order to have left the ``first-infall'' component at the current snapshot.
\item \textit{Post-pericentric CGM}: circum-galactic gas that has moved past pericenter of the orbit associated with first infall. 
This material can be outflowing (after pericenter) or inflowing (after apocenter). 
Diffuse gas that has been inside the CGM for more than a halo dynamical time will almost always belong to this component, 
unless it is identified as having been influenced by feedback, or was stripped from a satellite.
\item \textit{Feedback-influenced CGM}: circum-galactic gas that has been positively identified as having been likely influenced by stellar feedback in the past. 
This category overrides all of the other CGM classifications.
\end{itemize}

Gas outside the virial radius of the primary host halo is always labelled as ``first-infall'', unless
it is within the virial radius of another halo (in which case it is then labelled ``satellite gas''), or was ejected from the primary host.
Note therefore that when considering \lya surface brightness profiles that extend beyond the virial radius, we label
\lya photons as connected to ``satellites'' if they are either associated to a satellite subhalo of the primary, or if they are associated with
a neighbouring halo outside of the virial radius of the primary halo.

We estimate which tracers have been influenced by feedback (within the main halo) at a given snapshot by requiring that all of the following conditions are met:
\begin{itemize}
\item the gas is outflowing with a radial velocity (relative to the primary host halo center) of $v_{\mathrm{r}} > 50 \, \mathrm{km s^{-1}}$,
\item the gas has accelerated radially by $\Delta v_{\mathrm{r}} > 30 \, \mathrm{km s^{-1}}$ over the last $10 \, \mathrm{Myr}$,
and the gas has moved outwards by $\Delta r > 0.1 \, R_{\mathrm{vir}}$ over the same interval,
\item the radial kinetic plus thermal specific energy of the gas has increased by at least $25 \, \%$ over the same interval,
\item the tracer is not going to enter within the tidal radius of a satellite over the next $30 \, \mathrm{Myr}$. 
This can be important because the halo finder often loses track of satellites within the inner regions of the host halo,
\item the tracer has \textit{not} accelerated radially outwards simply due to moving past pericenter of its (gravity-driven) orbit.
\end{itemize}

We estimate if tracers are undergoing such an orbital transition at a given snapshot by first computing the 
minimum radial velocity (where inflowing is negative) of the tracer at previous snapshots, searching back
in time for as long as the tracer velocity is monotonically decreasing. For a particle that has just moved past pericenter,
this procedure identifies the maximum past infall velocity (and will return a smaller absolute value for particles that fluctuate
in velocity while settled within the ISM, corresponding roughly to the ISM velocity dispersion). 
We then search for the corresponding maximum radial
velocity of the tracer for future snapshots, searching forwards for as long as the radial velocity is monotonically increasing
with time. This then returns the maximum future outflowing velocity of the particle before it starts to slow
as it approaches apocenter. For particles that are on pure infalling orbits (independent of feedback or other
energy injection mechanisms), the maximum future outflow velocity will always be similar or less than the 
maximum past inflow velocity, and so for the tracer to be considered influenced by feedback we require that 
the maximum future outflow velocity be greater than $3/2$ times the magnitude of the maximum past infall velocity.

This scheme is of course approximate, and was designed and tested primarily by manual inspection of many individual tracer trajectories.
In practice, feedback events will affect many more of the tracers in the simulation at some level, and there
can be other physical mechanisms that might cause a tracer to suddenly accelerate outwards.

\subsection{\lya emission and radiative transfer}

With a full radiation hydrodynamics simulation, we self-consistently account for both the photo-ionization
and photo-heating of hydrogen by local stellar sources\footnote{Note that we do not
model supermassive black holes in our simulation, and so we do not account for photo-heating and ionization
by any local active galactic nuclei.} (plus a uniform UVB), and so we can compute the
\lya luminosity of all gas cells within the zoom-in region based on the relevant collisional excitation 
and recombination rates. 

For recombinations, we assume case B (see Blaizot et al., in prep for a full justification of this choice)
and follow \cite{Cantalupo08}, computing the rate of \lya photons produced by recombinations from a single
gas cell as

\begin{equation}
\varepsilon_{\mathrm{rec}} = n_{\mathrm{e}} \, n_{\mathrm{p}} \, \epsilon_{\mathrm{Ly\alpha}}^{B}(T) \, \alpha_{B}(T) \times (\Delta x)^3,
\end{equation}

\noindent where $n_{\mathrm{e}}$ and $n_{\mathrm{p}}$ are the electron and proton number densities,
$\alpha_{B}(T)$ is the case B recombination rate, $\epsilon_{\mathrm{Ly\alpha}}^{B}(T)$
is the fraction of recombinations that result in the production of a \lya photon, and $(\Delta x)^3$
is the volume of the cell. We evaluate $\epsilon_{\mathrm{Ly\alpha}}^{B}(T)$ using the fit from 
\cite[][given by their equation 2]{Cantalupo08}, and $\alpha_{B}(T)$ with the fit from 
\citet[][their appendix A]{Hui97}.

For collisional excitations, we compute the rate of resulting \lya photons produced from a single cell
as 

\begin{equation}
\varepsilon_{\mathrm{col}} = n_{\mathrm{e}} \, n_{\mathrm{HI}} \, C_{\mathrm{Ly\alpha}}(T) \times (\Delta x)^3,
\end{equation}

\noindent where $n_{\mathrm{HI}}$ is the number density of neutral hydrogen atoms, and $C_{\mathrm{Ly\alpha}}(T)$
is the rate of collisional excitations from level $1s$ to $2p$, which we evaluate using the fit from
\citet[][their equation 10]{Goerdt10}. In addition, we set the collisional excitation rate to zero
for cells within which the net cooling timescale is less than ten times the simulation timestep,
defining the net cooling timescale as $t_{\mathrm{cool,net}} \equiv (\mathcal{C}-\mathcal{H})^{-1}$, where $\mathcal{C}$ and $\mathcal{H}$
are the cooling and heating rates. In such cells, we do not adequately resolve the balance between
heating and cooling, which may lead to a significant overestimate or underestimate of the true
collisional excitation rate. Setting the collisional excitation rate to zero means that we are computing 
a conservative lower limit for these cells. We do however resolve the net cooling timescale in at least
$90 \, \%$ of the cells within the virial radius, and so we do not believe that this introduces a large uncertainty to the
collisional excitation contribution to \lya (Blaizot et al., in prep).

With the \lya emission rates from recombinations and collisional excitations, we then perform Monte Carlo 
radiative transfer in post-processing to compute the \lya signal that escapes the halo, using the 
\rascas code \cite[][]{MichelDansac20}. We sample each emission component 
($\varepsilon_{\mathrm{rec}}$ and $\varepsilon_{\mathrm{col}}$) with $4\times 10^5$ Monte Carlo photon 
packets cast from cells with a probability proportional to their luminosity. Each photon packet is emitted 
with a frequency in the frame of the emitting cell which is drawn from a Gaussian distribution of width 
fixed by the cell's temperature \citep[see Sec. 2.2 of][]{MichelDansac20}. In order to measure the 
extended emission out to large distances, all cells within a radius of $11 \, \mathrm{arcsec}$ around the main galaxy are 
allowed to emit. In turn, photon packets are propagated until they either pass that radius or are 
absorbed by a dust grain. We have included dust following the model of \citet{Laursen09b}, assuming SMC 
dust properties. In this model, the amount of dust scales primarily with the amount of neutral hydrogen 
and of metals, although traces of dust will be present in the ionised phase as well. We refer to 
Blaizot et al. (in prep.) for a detailed discussion of the effect of dust on the \lya properties of our 
simulated galaxy. This model results in UV attenuations 
(typically $\sim 1-2$ mag at $M_{1500} \sim -20$ AB) and \lya escape fractions (typically a few percent) 
which are broadly consistent with what is observed for LAEs. In the scope of the present paper, the main 
effect of dust is to limit the \lya flux emerging from the ISM of the central plus satellite galaxies.

In order to accelerate the computation, we use the core-skipping algorithm described 
in \citet[][Sect. 3.2.4]{Smith15}, which very slightly underestimates the effect of dust (introducing 
a relative error of order 1\% on the escape fraction of \lya radiation) but produces a speedup of the 
computation of a factor $\sim 1000$.
We do not apply any attenuation correction for the intergalactic medium, but this is generally expected
to have at most a modest $20-30 \, \%$ effect for wavelengths shorter than the line center for the redshifts 
($z \sim 3$) studied here \cite[e.g.][]{Inoue14,Hayes20}.

The main output of RASCAS is the state of photon packets as they escape the computational domain. With this 
data, we can perform part of our analysis and for example understand how different gas phases contribute to 
the observed extended emission. This does however limit our analysis to angle-averaged properties of the 
\lya emission, which to some extent inhibits our ability to compare our simulation directly to MUSE observations. 
In Appendix~\ref{ap_11_comp}, we make use of the peeling algorithm \citep{YusefZadeh84,Zheng02}, as described 
in \citet[][Sect. 9.3]{Dijkstra17}, to collect flux in 12 mock MUSE data-cubes at each snapshot. 
These 12 mocks correspond to observations of the simulated galaxy from 12 different directions (approximately
equally spaced, with a common origin, and that are the same for all redshift outputs), allowing us to more closely reproduce the observational selection and analysis of \cite{Wisotzki18}.

\section{Results}
\label{flows_sec}

Our results are presented as follows: we start by using tracer particles to explore the origin of the
dense gas phases of the CGM in Section~\ref{lag_orig}, we connect this to the properties of \lya emission
related to the CGM in Section~\ref{composition_subsec}, we decompose the different components
that comprise observed \lya haloes in Section~\ref{nature_obs_haloes}, we compare \lya
surface brightness profiles with stacked MUSE observations in Section~\ref{MUSE_comp_sec}, and
we finish by presenting the predicted \lya spectral morphology in Section~\ref{spectral_morphology_sec}.
Since our focus is primarily directed at the connection between \lya emission and the circum-galactic
medium, we generally exclude gas (and photons) associated with satellite galaxies for the analysis presented in 
Sections~\ref{lag_orig} and \ref{composition_subsec}. We then introduce the contribution of satellites
when considering the observed properties of our simulated galaxy, in Sections~\ref{nature_obs_haloes},~\ref{MUSE_comp_sec} and ~\ref{spectral_morphology_sec}.

\subsection{Lagrangian origin of the neutral hydrogen in the CGM}
\label{lag_orig}

\begin{figure}
\includegraphics[width=20pc]{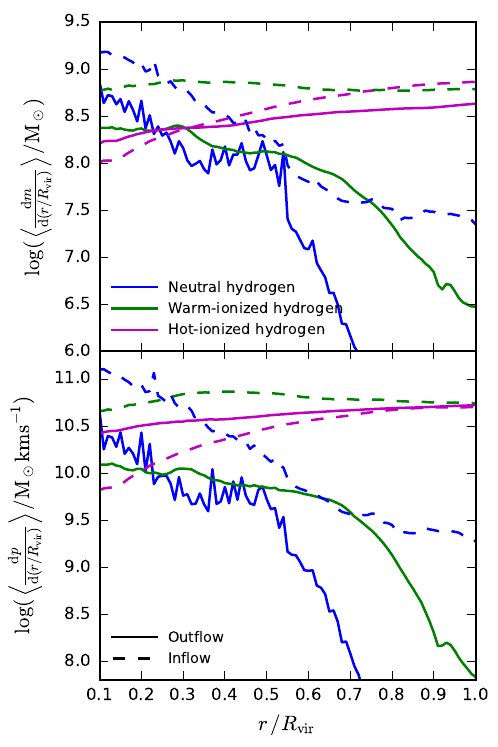}
\caption{Mass and radial momentum profiles of circum-galactic gas (excluding satellite galaxies), 
plotted as a function of radius.
Data is mean-stacked over all available outputs between $z=4$ and $z=3$.
Solid lines show outflowing gas, and dashed lines show inflowing gas.
Blue shows neutral hydrogen (i.e. the mass/momentum of hydrogen in each cell
multiplied by the neutral hydrogen fraction), green shows warm-ionized hydrogen ($T<10^{4.5} \,\mathrm{K}$), 
and magenta lines show hot-ionized hydrogen ($T>10^{4.5} \,\mathrm{K}$).
Most of the neutral hydrogen content of the CGM is 
located within the inner halo, whereas the hot-ionized profile rises with radius, 
comprising the majority of circum-galactic gas in the outer halo.
The sharp drop in neutral hydrogen mass/momentum with radius could be associated with a galactic fountain
process, or with a change in ionization state of gas as it flows inwards/outwards.
}
\label{mp_profiles}
\end{figure}

In \cite{Mitchell18}, we analysed the properties of circum-galactic gas from a sample of zoom-in simulations
at $z \approx 3$, including the halo studied here. Focussing on the neutral phase of hydrogen (which we expect
to be the closely related to the gas phases associated with extended \lya emission), we found that the mass per 
unit radius in neutral hydrogen\footnote{By neutral neutral hydrogen mass we mean $0.76 \, x_{\mathrm{HI}} \, \rho_{\mathrm{gas}} \, \Delta x^3$,
where $0.76$ is the assumed primordial hydrogen mass fraction, and $x_{\mathrm{HI}}$ is the hydrogen neutral fraction
of a given cell.}
drops exponentially as a function of radius. In contrast, the mass profile of warm and hot ionized CGM components 
is generally flat or rising with radius. This is shown for the simulation studied here in the top panel
of Fig.~\ref{mp_profiles}. Most of the neutral hydrogen is radially inflowing (dashed blue line), but a significant
fraction ($\approx 1/3$) is outflowing (solid blue line) in the inner regions ($r < 0.5 \, R_{\mathrm{vir}}$).
Under the approximation of constant radial flow velocity, a drop in the neutral hydrogen mass with radius
implies that mass is being lost from the phase as gas flows outward, and correspondingly that the inflowing
neutral component is gaining mass as gas flows inwards. This is confirmed by the lower panel of Fig.~\ref{mp_profiles},
which shows similar gradients for the radial momentum profiles.

Understanding this behaviour is an important prerequisite for understanding in turn the predicted spatially extended \lya
signal. Our simulation predicts copious amounts of outflowing neutral hydrogen in the 
inner CGM, which may be consistent with the picture of outflowing neutral gas shells that is commonly invoked
to explain the \lya spectral morphology of observed LAEs \cite[e.g.][]{Ahn04,Verhamme06}. At larger radii 
($r>0.5 \, R_{\mathrm{vir}}$), there is comparatively very little neutral hydrogen in our simulation,
which poses the question of whether the outflowing neutral gas recorded in the inner CGM is being
photo-ionized as it moves outwards (joining the warm-ionized component, solid green line, which is more
radially extended than the neutral outflowing component), or alternatively is falling back towards the ISM in a fountain flow (before
reaching the outer CGM). Accordingly, the increase in inflowing neutral gas in the inner CGM could be attributed
to a change in ionization state of gas that is infalling for the first time, or could be connected
to the afore-mentioned fountain flow of neutral gas. 
Another question that arises is whether the neutral hydrogen
content of the CGM in the simulation is being strongly affected by feedback.
In \cite{Mitchell18}, we show that not including stellar feedback in our simulations approximately leaves 
the neutral content of the CGM unchanged,  implying that neutral circum-galactic gas flows in the simulation
are driven primarily by other processes.

\begin{figure*}
\begin{center}
\includegraphics[width=40pc]{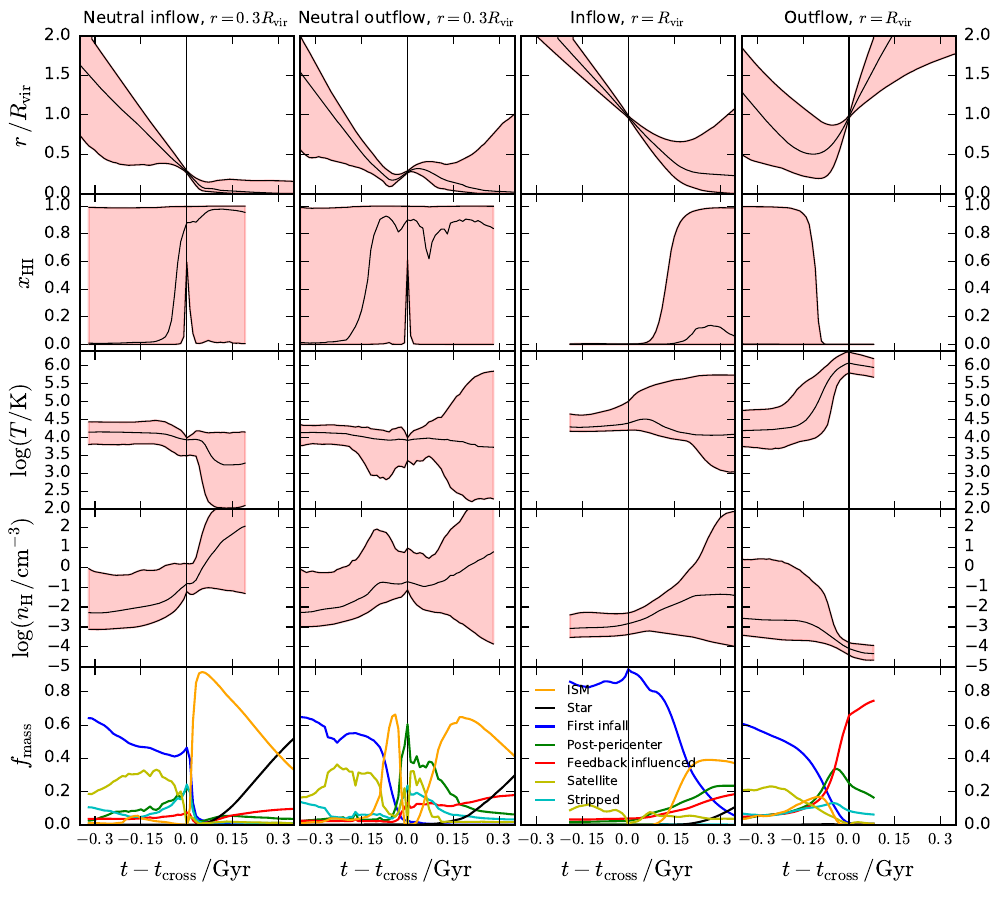}
\caption{An overview of the Lagrangian history of circum-galactic gas in the simulation.
Each panel shows weighted distributions of tracer trajectories, for tracers that meet sets of selection criteria.
Each column corresponds to a different set of selection criteria, as indicated by the column titles.
Selections are performed either at $0.25 < r/R_{\mathrm{vir}} < 0.3$, or at $0.95 < r/R_{\mathrm{vir}} < 1$,
and include either inflowing or outflowing gas.
Individual tracers that fulfil these criteria are shifted along the x-axis such that they pass 
the selection at $t = t_{\mathrm{cross}}$.
Tracer distributions are flux-weighted either by the total radial momentum at the time of selection (right-side columns), 
or by the radial momentum of neutral hydrogen at the time of selection (left-side columns).
Solid black lines indicate the $16^{\mathrm{th}}$, $50^{\mathrm{th}}$, and $84^{\mathrm{th}}$ percentiles of the distributions.
Each row corresponds to a different gas quantity, including radius, neutral hydrogen fraction, temperature, density, and 
the mass fraction, $f_{\mathrm{mass}}$, 
belonging to the discrete and mutually exclusive Lagrangian components described in \protect{Section~\ref{tracer_methods}}.
Lagrangian components include gas that is infalling for the first time (blue), 
that has been stripped from a satellite (cyan), that is within the tidal radius of a satellite (cyan),
that has been influenced by feedback (red),
that has moved past the first pericentric passage (green),
that belongs to the central ISM (orange),
that belongs to a satellite (khaki),
and tracers that are locked inside stars (black).
Distributions of gas temperature, density, and ionization fraction are truncated once $20 \, \%$ of the selected gas 
has either been locked into stars (first, second columns), 
or is located beyond $2 \, R_{\mathrm{vir}}$ (third, fourth columns), 
at which we point we stop tracking trajectories.
}
\label{fate_neut_outflows}
\end{center}
\end{figure*}

We can now address these questions directly by using the Lagrangian information afforded by tracer particles. 
Fig.~\ref{fate_neut_outflows} shows the Lagrangian history of gas selected as it crosses either
the virial radius, or a surface at $0.3 \, R_{\mathrm{vir}}$. We temporally stack simulation outputs around
the crossing redshift, such that the trajectory distributions are representative for this halo over $3.5<z<4$.
Focussing first on inflowing neutral hydrogen crossing $0.3 \, R_{\mathrm{vir}}$ (far-left column), the distribution
of tracer trajectories shows that most of the selected gas is flowing inwards for the first time from the wider
environment, as opposed to being recycled in a fountain flow. The temperature of the
inflowing gas ($T \sim 10^4 \, \mathrm{K}$, middle row) is fairly steady as the gas flows inwards down to 
$r \sim 0.5 \, R_{\mathrm{vir}}$, consistent with it being maintained in thermal equilibrium by 
photo-heating. At the same time, the density of the gas gradually increases (fourth row) until the gas reaches
the threshold to be self-shielded from ionizing radiation at $n_{\mathrm{H}} \sim 10^{-2} \, \mathrm{cm}^{-3}$, at which
point the neutral hydrogen fraction rapidly increases (second row).
From this, we conclude that the exponential increase in the mass/momentum of inflowing neutral hydrogen with decreasing
radius seen in Fig.~\ref{mp_profiles} is caused primarily by compression of infalling gas leading to a change of ionization
state\footnote{Note however that the distributions shown in Fig.~\ref{fate_neut_outflows} are weighted by flux, which
will somewhat down-weight (relative to weighting by mass) the importance of a neutral fountain flow that is turning around at a radius comparable to $0.3 \, R_{\mathrm{vir}}$.
Fig.~\ref{mp_profiles} does indeed show a jump in both the inflowing and (more prominently) the outflowing
neutral hydrogen radial mass profiles at $r \approx \, 0.5 R_{\mathrm{vir}}$, implying that fountain flows
do indeed also play a role.}.

Shifting our attention to the neutral \emph{outflowing} hydrogen crossing $0.3 \, R_{\mathrm{vir}}$ (second-left column 
in Fig.~\ref{fate_neut_outflows}), the past distribution of trajectories shows that this gas is comprised primarily
of infalling gas that has moved past first pericenter. The majority of this outflowing gas will soon move past apocentre of its
first orbit and will subsequently settle down towards the central ISM. The median average
ionization fraction stays at $x_{\mathrm{HI}} \sim 0.8$ after the selection redshift, meaning
that the sharp drop in outflowing hydrogen mass with increasing radius is primarily connected to
gas falling back down onto the ISM, rather than because of a change in ionization state.
Employing the Lagrangian tracer classifications described in Section~\ref{tracer_methods}, this component of the
gas is, as expected, associated with the ``post-pericenter'' component (green line in the bottom row), which comprises the majority
of the neutral outflow at $0.3 R_{\mathrm{vir}}$.
At the same time, a subset of the gas does move out into the outer CGM, reflected by the upper ($84^{\mathrm{th}}$) percentile in 
the radius distribution (top-row).  This is associated with a growth in the 
``feedback-influenced'' component (red line in the bottom row), implying that feedback is responsible for pushing 
this fraction of the gas out into the outer halo. 

The third and fourth columns of Fig.~\ref{fate_neut_outflows} give a broader perspective by looking at the trajectories of gas
before/after crossing the halo virial radius (in this case without selecting any specific gas phase). 
Starting with inflowing gas at the virial radius (third column), the future trajectories of the gas are diverse,
with $\approx 40 \, \%$ of the gas settling into the central ISM (orange line, bottom row), $\approx 20 \, \%$ moving past pericenter
but remaining in the inner CGM (green line, bottom row), and $\approx 20 \, \%$ moving back out of the halo, which
is connected with the component that is estimated to have been influenced by feedback (red line, bottom row).
This diversity is reflected by a broad range in temperature and density, spanning three orders of magnitude
in temperature and seven orders of magnitude in density (at $300 \, \mathrm{Myr}$ after entering the halo).
It is worth noting that the inflowing gas is almost entirely ionized at the virial radius, and on average remains mostly
ionized for at least $300 \, \mathrm{Myr}$ after entering the halo (so for at least one to two halo dynamical times).

\begin{figure*}
\begin{center}
\includegraphics[width=40pc]{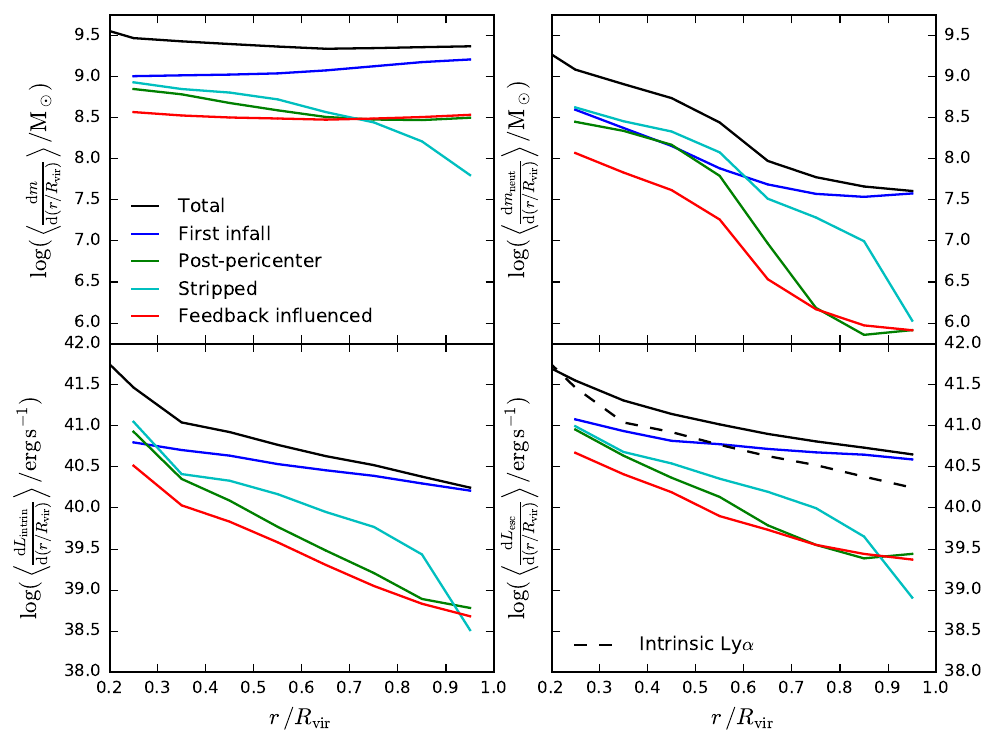}
\caption{Radial profiles of total gas mass (upper-left), neutral hydrogen mass (upper-right), 
and both intrinsic (lower-left) and escaping \lya luminosity (lower-right) for circum-galactic gas
(excluding satellite galaxies).
Unlike in \protect Fig.~\ref{mp_profiles}, here the radial profiles are split between different Lagrangian CGM components,
as indicated by the legend.
Data is mean-stacked over all available outputs between $z=4$ and $z=3$.
The intrinsic \lya luminosity profile (lower-left) shows the luminosity of emitting gas, irrespective of whether photons escape the halo.
The escaping \lya luminosity profile (lower-right) shows how escaping \lya photons last trace the CGM before escape, meaning that photons are
binned at the radial position of last-scattering (or at the emission location if no scattering occurs), and are assigned to the CGM component
associated with that position.
The dashed black line in the lower-right panel shows the total intrinsic \lya luminosity profile from the lower-left panel, 
for comparison.
Note that these the profiles are binned as a function of three-dimensional radius, and so the \lya profiles are not directly comparable to observed surface brightness profiles.
\lya emission is dominated by first-infalling gas in the outer CGM, both in terms of intrinsic luminosity and as the component being
traced prior to escape. \lya emission in the inner CGM traces gas associated with a mix of components.
Radiative transfer effects (i.e. scattering) enhances the escaping \lya profile relative to intrinsic by around a factor two in the CGM.
}
\label{cgm_profiles_lagrangian}
\end{center}
\end{figure*}

Considering instead outflowing gas at the virial radius (far-right column), this selection is dominated by the feedback
influenced component (red line, bottom row). The majority of this gas did not come from the ISM, but was instead accelerated
outwards while infalling (for the first time) within the CGM, with an associated increase in temperature from $T \sim 10^{4} \, \mathrm{K}$
to $10^6 \, \mathrm{K}$, consistent with being heated and entrained by hot and diffuse supernova-driven outflows.
Half of this component did not even reach $r<0.5 \, R_{\mathrm{vir}}$ before being accelerated outwards.

\subsection{The composition of the high-redshift \lya-emitting CGM}
\label{composition_subsec}

Combined together, Fig.~\ref{mp_profiles} and Fig.~\ref{fate_neut_outflows} present a picture in which most of the 
the neutral (and so higher-density) phase of the CGM is undergoing gravitational infall, with the majority of 
neutral inflows comprised of gas infalling before first-pericentric passage, and the majority of neutral outflowing
gas comprised of gas that has moved past first pericenter, and is in the process of settling into an equilibrium
configuration either within the inner CGM, or down in the dense ISM at the halo centre.

\subsubsection{Radial profiles}

\begin{figure*}
\begin{center}
\includegraphics[width=40pc]{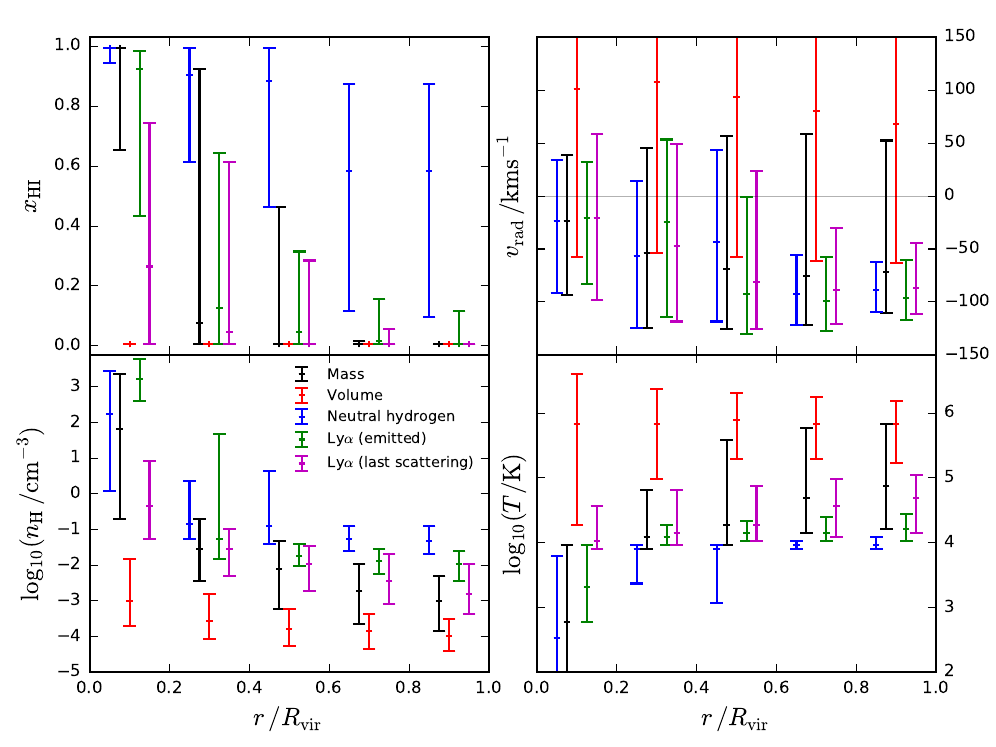}
\caption{Distributions of neutral hydrogen fraction ($x_{\mathrm{HI}}$, top-left),
radial velocity ($v_{\mathrm{rad}}$, top-right),
hydrogen number density ($n_{\mathrm{H}}$, bottom-left), 
and gas temperature ($T$, bottom-right), all plotted as a function of radius.
Gas within satellite galaxies is excluded.
Points indicate the $16$, $50$, and $84^{\mathrm{th}}$ percentiles of distributions weighted
either by mass (black), volume (red), neutral hydrogen mass (blue), 
intrinsic \lya luminosity (green), 
or the luminosity of \lya photons that last interact with a given gas cell before escaping (magenta).
The vast majority of escaping photons scatter at least once before escaping.
Different colour points are horizontally offset from each other for plotting purposes only, and are
all computed from a common set of five radial bins, ranging from $r=0$ to $r=R_{\mathrm{vir}}$.
Generally speaking, \lya photons are emitted and (last) scatter from partially (or near fully)
ionized and inflowing gas that is heated to slightly above $10^4 \, \mathrm{K}$.
In the CGM, \lya generally traces gas that is denser than the average (defined either by mass or volume),
but that is less dense than the neutral hydrogen component of the CGM.
}
\label{phase_profiles}
\end{center}
\end{figure*}

This picture is shown explicitly in Fig.~\ref{cgm_profiles_lagrangian}, which splits the total CGM (top-left panel)
and neutral hydrogen content of the CGM (top-right) between the various Lagrangian components described in 
Section~\ref{tracer_methods}. The total gas content (top-left panel) of the outer CGM is dominated by the inflowing
gas that is infalling for the first time (``first infall'', blue line), with the contribution from gas that has moved past first
pericentric passage (``post pericenter'', green line) gradually increasing until it becomes comparable to the first-infall
component at $r \sim 0.2 \, R_{\mathrm{vir}}$. Gas that has been stripped/removed from satellites (``stripped'', cyan line) 
is also an important contributor in the inner CGM (and in practice is comprised of both gas that is infalling for the
first time, and gas that has moved its first pericentric passage). Gas that has been estimated to have been clearly 
influenced by stellar feedback (``feedback influenced'', red line) is subdominant at all radii, but with a flat radial 
mass profile, corresponding to the approximately flat mass and momentum profiles seen for the hot-ionized outflowing phase
in Fig.~\ref{mp_profiles}.

Shifting attention to the neutral hydrogen content of the CGM (top-right panel in Fig.~\ref{cgm_profiles_lagrangian}),
the drop in neutral hydrogen mass with radius is evident for each individual Lagrangian component.
The relative contribution of stripped gas is higher for the neutral phase, and there is a greater level of parity between
the first infall, post pericenter, and stripped components, particularly within $r < 0.6 \, R_{\mathrm{vir}}$.
At larger radii, the first-infalling component increasingly dominates the neutral gas profile, although the
vast majority of the hydrogen at $R_{\mathrm{vir}}$ is in an ionized phase. 

The bottom panels of Fig.~\ref{cgm_profiles_lagrangian} show the connection with \lya emission,
splitting the luminosity profile between the same Lagrangian components.
Interestingly, the intrinsic \lya luminosity emitted in-situ within the CGM (lower-left panel) appears to scale with radius
in a manner that is between the total and neutral hydrogen mass profiles. 
As with the neutral hydrogen mass profile,
the intrinsic \lya luminosity decreases with radius for each component. The \lya luminosity declines less strongly
with radius however, and the relative contribution from first-infalling gas is more reminiscent
of the total mass profile.  This implies that the intrinsic \lya emission is not being
dominated by the neutral phase (despite the higher associated densities), and a significant contribution
to the total intrinsic \lya luminosity of the CGM comes from warmer ionized gas, particularly in the outer CGM.
Note that satellite galaxies are excluded from our analysis at this stage, which would otherwise dominate
the intrinsic \lya luminosity outside of the central ISM (but not the escaping luminosity).

The lower-right panel of Fig.~\ref{cgm_profiles_lagrangian} focuses on escaping \lya emission as a
tracer of the CGM. The solid black line shows the total \lya luminosity that escapes the halo after last
scattering from a given radius within the CGM (or from the radius of emission if the escaping photons do not scatter).
This (unlike the intrinsic CGM emission profile) therefore does include a contribution from photons that are emitted within the central ISM or within
satellite galaxies, but only if those photons subsequently scatter off of circum-galactic gas.
As with the other panels, the coloured lines then indicate the contribution of photons that last scatter
off of different CGM components. In general, the escaping \lya photons trace the CGM in much the same
way as the intrinsically emitted photons shown in the lower-left panel.
As a reference, the dashed black line shows the intrinsic \lya luminosity of the CGM, showing
that radiative transfer effects (i.e. scattering of photons emitted in the central ISM or 
within satellites) boost the escaping \lya signal tracing the CGM out to the virial radius ($\sim 30 \, \mathrm{kpc}$)
by roughly a factor two relative to the intrinsic profile. 

The gas phases traced by \lya emission are shown directly in Fig.~\ref{phase_profiles}, in which
distributions of gas properties are plotted, weighted by mass, volume, or associated \lya luminosity.
As was indicated by the shape of the radial mass/luminosity profiles shown in 
Fig.~\ref{cgm_profiles_lagrangian}, \lya generally traces partially (or fully) ionized gas that
is heated to slightly above $10^4 \, \mathrm{K}$, rather than tracing the fully neutral phase
directly. \lya does still trace gas that is generally denser than the average within the CGM, defined
either by total mass or volume weighting. Escaping \lya photons (magenta points) last-scatter off gas
that is warmer and lower in density than the \lya emitting gas (green points) in the CGM.
The vast majority of escaping photons do scatter at least once.
Notably, within the ISM (far-left radial bin) escaping \lya photons last-scatter from warm, partially ionized gas (median
$x_{\mathrm{HI}} \approx 0.2$) that sits at densities above the imposed threshold for self-shielding 
from the UVB ($n_{\mathrm{H}} = 0.01 \, \mathrm{cm^{-3}}$), highlighting the impact of photo-heating and photo-ionization from local radiation sources.

Putting the information in Fig.~\ref{cgm_profiles_lagrangian} and Fig.~\ref{phase_profiles} together,
we conclude that \lya is a fairly faithful differential tracer of mass in the different Lagrangian components that we
consider, albeit with declining luminosity as a function of radius due to the associated drop in average gas densities
at larger radii. That said, our analysis so far has only considered both time and angle-averaged quantities,
and in principle \lya may trace specific components of the CGM preferentially if viewed from a specific direction
or at a given time. This is pertinent, given that only a subset of high-redshift galaxies are observed to be LAEs, 
and that at least for now there is an open question in observations as to whether all rest-frame UV-visible high-redshift 
galaxies have associated spatially extended \lya haloes.


\subsubsection{Temporal variations}

\begin{figure*}
\begin{center}
\includegraphics[width=40pc]{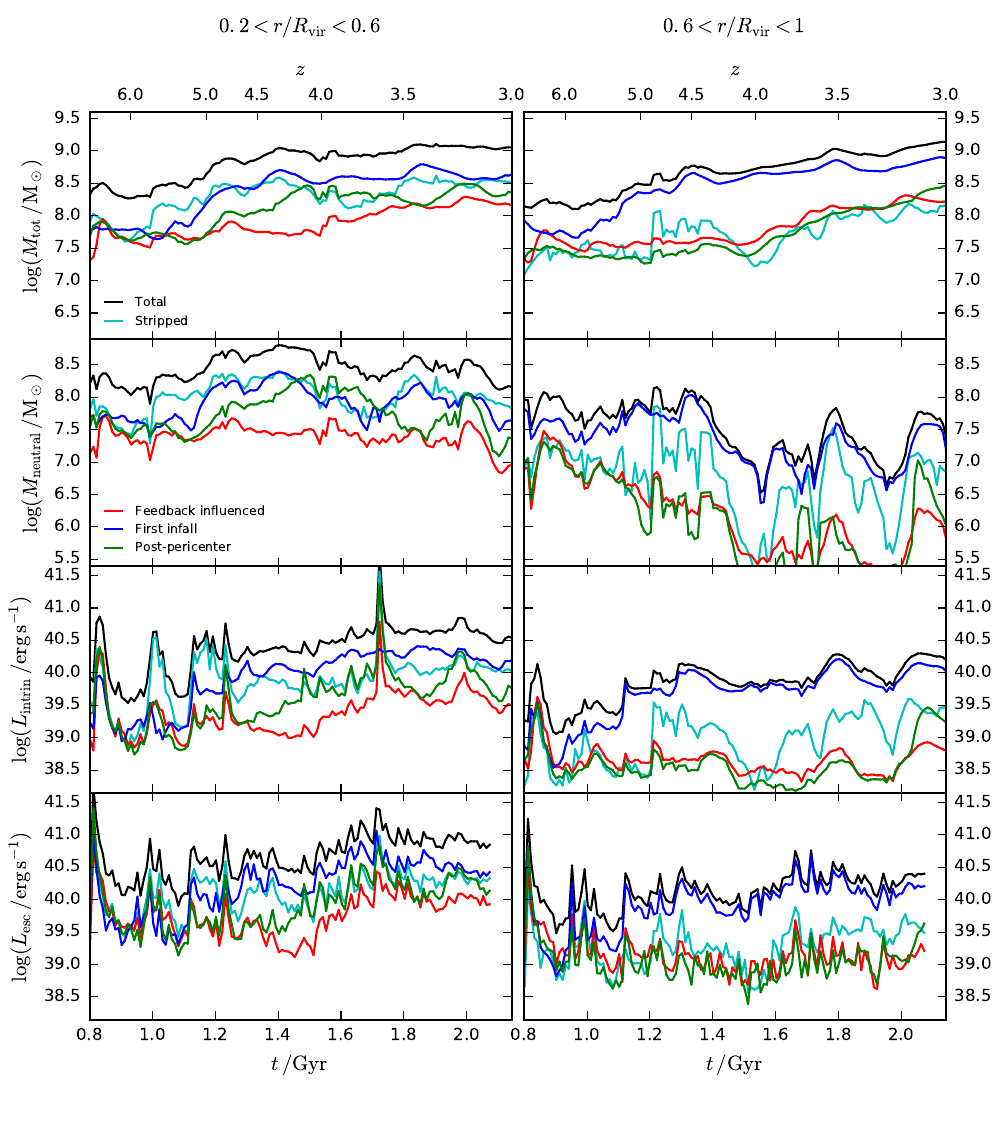}
\caption{Time evolution in mass/luminosity of different CGM components (excluding satellite galaxies).
Different rows show total gas mass (top row), neutral hydrogen mass (second row), 
intrinsic \lya luminosity (third row), and escaping \lya luminosity (bottom).
The left column shows the inner CGM over $ 0.2<r \, / R_{\mathrm{vir}} < 0.6 $,
and the right column shows the outer CGM over $ 0.6<r \, / R_{\mathrm{vir}} < 1 $.
Generally speaking, temporal fluctuations are relatively modest after $z=5$; with 
escaping \lya luminosity tracing the CGM only varying by factors of a few over short timescales. 
}
\label{cgm_origin}
\end{center}
\end{figure*}

Fig.~\ref{cgm_origin} shows the time dependence of different
spatially integrated CGM properties for both the inner CGM ($0.2<r\,/R_{\mathrm{vir}}<0.6$, left column) and the outer
CGM ($0.6<r\,/R_{\mathrm{vir}}<1$, right column). The total mass in the CGM (black line, top row) is
approximately constant for $z<4.5$, and generally speaking this applies to the individual Lagrangian components, with
only factors of a few variations over timescales of around $100 \, \mathrm{Myr}$.
Neutral hydrogen in the CGM varies more strongly in time (second row), particularly in the outer CGM where order
of magnitude variations are seen over $100 \, \mathrm{Myr}$ timescales. The contribution from gas stripped (or
otherwise removed) from satellites varies particularly strongly, as individual massive satellites can dominate
the signal.

Strong time variations in the intrinsic \lya luminosity of the inner CGM (third row, left column) are seen at high 
redshift ($z>5$), but subsequently variations are fairly modest, with the exception of a strong feature at $z \sim 3.6$.
From visual inspection this appears to be caused by dense star-forming ISM gas briefly moving outside of $0.2 \, R_{\mathrm{vir}}$
during a satellite merger. Considering also the outer CGM, the temporal variations in intrinsic \lya luminosity are
modest at later times, particularly when compared to the mass in neutral hydrogen.

The bottom row of Fig.~\ref{cgm_origin} shows the last CGM component that \lya photons interacted with
before escaping the halo. The total escaping \lya luminosity that traces the CGM varies only modestly in time. Short term
fluctuations of about a factor two are apparent, but are partly driven by the Poisson noise in the finite distribution
of \lya photons used to perform Monte Carlo radiative transfer. 
We conclude that temporal variations are not a dominant leading-order effect for \lya emission from the CGM in our simulation, 
at least when averaged over all angles. This has the implication that high-redshift galaxies should generally
be surrounded by a diffuse \lya halo, irrespective of the brightness of the central \lya component (which
does vary strongly in time, Blaizot et al., in preparation), or even if there is no central emission
component (i.e. net \lya absorption against the continuum).

\subsection{The nature of observed \lya haloes}
\label{nature_obs_haloes}

\begin{figure}
\includegraphics[width=20pc]{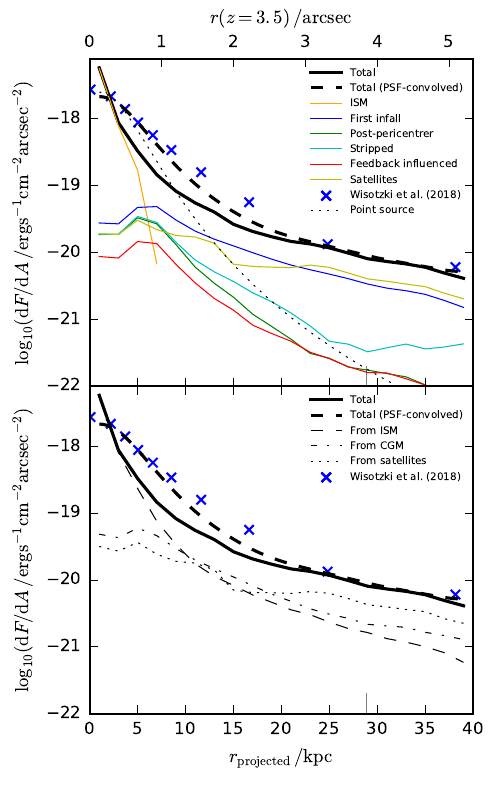}
\caption{Mean stacked ($3<z<4$) and angle-averaged \lya surface brightness profiles, 
plotted as a function of projected separation from the halo center.
The thick solid black line in both panels shows the total escaping \lya surface brightness profile from the simulation.
The thick dashed black line shows the same profile after convolving with the typical 
point spread function (PSF) inferred for deep-field MUSE observations.
For reference, the dotted line in the top panel shows the PSF-convolved profile of a scaled point source.
The virial radius of the simulated halo at $z=3.5$ is $R_{\mathrm{vir}} = 29 \, \mathrm{kpc}$, 
as marked by a small grey indicator at the bottom of each panel.
In the top panel, thin coloured lines show how escaping \lya emission traces different components of the CGM,
with each photon associated with the component it last interacted with before escape.
For the projected profiles shown here, we now also include photons that escape 
from the central ISM (orange line), and from satellite galaxies 
(khaki line), without interacting with the intervening CGM. 
In the bottom panel, thin lines show the origin of escaping \lya emission, indicating whether
escaping photons at a given projected radius were originally emitted within 
the central ISM (dashed line), in situ within the CGM (dash-dotted),
or within a satellite galaxy (dotted).
For comparison, the stacked \lya surface brightness profile of $92$ detected \lya emitters from the MUSE HDFS and UDF fields 
is shown by blue crosses, taken from \protect \cite{Wisotzki18}.
}
\label{SB_profile}
\end{figure}

\begin{figure}
\includegraphics[width=20pc]{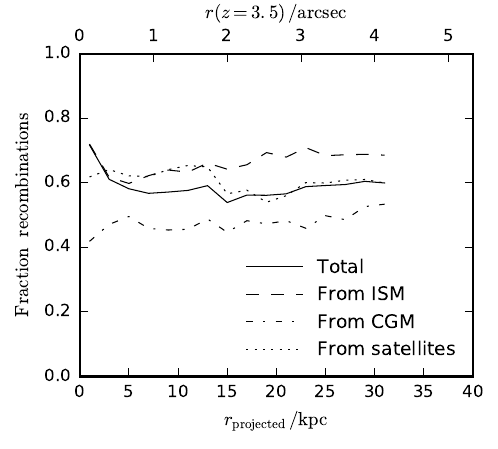}
\caption{The fraction of the projected \lya surface brightness contributed by photons produced in recombinations
(as opposed to collisional excitations), plotted as a function of projected separation from the halo center.
As in \protect Fig.~\ref{SB_profile}, data are taken from a mean, angle-averaged stack of our simulated galaxy over $3<z<4$
(though note that no PSF convolution is applied here).
Lines show the fractional contribution of recombinations for the total surface brightness profile (solid line),
and for photons that were originally emitted inside the ISM (dashed), inside satellite galaxies (dotted), and in-situ within
the CGM (dash-dotted). Collisional excitations provide a fractionally larger contribution to the escaping
\lya flux produced in-situ within the CGM ($\approx 50 \%$), whereas recombinations contribute the majority
of the escaping flux for photons emitted within the ISM ($\approx 65 \%$). 
Recombinations generally contribute $\approx 60 \%$ of the total escaping flux.
}
\label{rec_col_split}
\end{figure}

Fig.~\ref{SB_profile} shows radial surface brightness profiles of escaping \lya photons, 
now plotted as a function of projected circular radius (relative to the direction of each
escaping photon). The profile is mean stacked over redshifts $3<z<4$, and over all
possible viewing angles.
In the top panel, we split the escaping \lya photons based on the Lagrangian CGM 
component that each photon last interacts with before escape, using the Lagrangian components 
described in Section~\ref{tracer_methods}. In addition to the CGM components discussed
previously, we now also show the contribution from photons that 
last scattered on the central ISM (orange lines, defined here as 
$r < 0.2 \, R_{\mathrm{vir}}$, where $r$ in this case is the three-dimensional radius), or 
last-scattered within satellites (khaki lines), without interacting with the intervening CGM.

While it is expected that the central ISM dominates the production of photons observed near 
the centre of the projected profile, it is notable that the majority of the photons observed 
in the central region of the profile ($r_{\mathrm{projected}} < 7 \, \mathrm{kpc}$) do not 
scatter in the CGM (orange line). At larger separations, the two main contributors to the signal 
are the ``first infall'' gaseous component (blue line) and the component of photons
that escape from satellite galaxies (khaki line).
If the simulation picture is correct, this implies that observed \lya haloes are 
tracing primarily inflowing circum-galactic gas along with emission escaping
from satellites (at least when stacked).

The lower panel of Fig.~\ref{SB_profile} splits the escaping \lya emission based on the 
original source of the emission (i.e. before scattering), splitting between the ISM of the central galaxy (dashed lines), the ISM of 
satellites (dotted), and in-situ photons emitted within the CGM (dash-dotted). The 
largest contributor at projected separations $< 10 \, \mathrm{kpc}$ is from photons that 
were emitted within the ISM of the central galaxy (note that this is before any point spread function
convolution is applied). At larger separations the escaping radiation is comprised of
a roughly equal mix of scattered photons from the central ISM, in-situ CGM photons, and photons emitted
within satellites\footnote{As a reminder, we refer to both satellite subhaloes (within the virial
radius of the primary host halo) and neighbouring haloes (outside the virial radius of the primary host halo)
as ``satellites''.}. 
Over $10 < r_{\mathrm{projected}} \, / \mathrm{kpc} < 20$ the in-situ
component provides marginally the largest contribution, with satellites increasingly powering the
profile at larger separations \cite[forming an effective ``two-halo term'' outside the halo 
virial radius, see also][]{Zheng11, MasRibas16, MasRibas17}, though at these radii
the stacked surface brightness has dropped significantly below the $\sim 10^{-19} \, \mathrm{erg s^{-1} cm^{-2} arcsec^{-2}}$
limit of non-stacked MUSE deep-field observations of individual LAEs \cite[e.g.][]{Wisotzki16,Leclercq17}.
The transition with projected separation from central-dominated, to diffuse CGM, to faint satellites setting the SB profile is consistent
with the simulations-based study of \cite{Lake15}, who simulate a brighter LAE at the same redshifts.
Note also that after convolving with the MUSE PSF, the central ISM component is pushed outwards,
and contributes comparably to the other components out to $\sim 25 \, \mathrm{kpc}$ in projection.

An important outstanding question for the interpretation of observed \lya haloes is the nature of the
powering mechanism for the emission, which can be scattering of \lya photons escaping the ISM, emission associated with
satellite galaxies (which are likely unresolved even in deep observations of LAEs), in-situ
emission powered by ionizing UV radiation escaping the galaxy or from the UVB, or alternatively
in-situ emission powered by compressive heating associated with gravitational infall or feedback processes. While
we have addressed the role of scattering and satellite galaxies here, the powering of the in-situ
CGM emission is not trivially known even in a simulation. It is sometimes
assumed that collisional excitations in the CGM are predominantly powered by gravitational heating,
and that recombinations in the CGM are powered by fluorescence, but in practice compressive heating
of dense gas will lead to recombinations (though admittedly we expect photo-ionization to be the dominant
ionization mechanism) and photo-heating from UV photons will lead to collisional
excitations. 

We leave the task of separating these effects for future work, but we do consider the
relative contribution of recombinations versus collisional excitations in Fig.~\ref{rec_col_split}.
Recombinations provide the majority of escaping photons that were emitted within the ISM
(about $65 \, \%$), but collisional excitations provide around $50 \, \%$ for the in-situ
CGM component. This implies that different physical processes are likely powering
the in-situ CGM emission relative to the ISM emission.

\subsection{Comparison to surface brightness profiles from stacked MUSE observations}
\label{MUSE_comp_sec}

Fig.~\ref{SB_profile} also compares our stacked simulation profile with the stacked
\lya surface brightness profiles measured using MUSE observations of $92$ $3<z<4$ LAEs from the Hubble Deep 
field South, and the Hubble Ultra Deep Field \cite[][]{Wisotzki18}. The thick solid-dashed
line in Fig.~\ref{SB_profile} shows our simulation prediction after convolving with
the appropriate point spread function (PSF). We adopt a Moffat profile with 
$\beta = 2.8$, and $\mathrm{FWHM} = 0.875 - \frac{1}{3}(\lambda \, / 10^4 \,\mathrm{\AA})$,
as inferred for the relevant deep-field MUSE observations \cite[][]{Bacon17}.
For reference, we also show the profile of a scaled point source, convolved with
the same PSF. As discussed in \cite{Wisotzki16,Wisotzki18}, MUSE clearly resolves
spatially extended emission outside of the central projected arcsecond.

After PSF convolution the agreement between the observed and simulated stacked profiles 
is generally excellent, though perhaps partially fortuitous given that we have 
only simulated a single galaxy. The agreement in shape within the inner $\sim 7 \, \mathrm{kpc}$
in projection is not necessarily physically meaningful, given that the profile on these
scales is dominated by PSF-broadening (comparison to the dotted line, top panel).
For larger scales where PSF-broadening is sub-dominant, the simulated profile under-predicts the observed
profile by up to $50 \, \%$ for the range $10 < r_{\mathrm{projected}} \, / \mathrm{kpc} < 20$,
which may be significant given that this is the range where we predict in-situ CGM emission
to play the largest role. This could be interpreted as evidence that we have insufficient
dense and/or neutral gas in the CGM of our simulated halo, possibly because our simulated galaxy is
too low in halo mass (we believe our simulated galaxy has a stellar mass which is 
too large relative to its halo mass, see discussion in Section~\ref{mhalo_problem}),
or possibly because we are missing a dense/neutral phase associated with supernova-driven outflows.
For yet larger separations ($r_{\mathrm{projected}} \, / \mathrm{kpc} > 20$) close
to and beyond the halo virial radius, agreement with the observations is excellent.
Overall, the level of agreement with observations is encouraging, and inspires a level of confidence
in our results regarding the origin of observed \lya haloes (though see the forthcoming
discussing on \lya spectral morphologies).

Note that the simulated surface brightness profile does not exactly follow the 
observational selection and stacking procedure. Specifically, the
simulated profile shown here is mean stacked and uses angle-averaged \lya
outputs. This is necessary in order to split photons into different components\footnote{
Median stacking requires the production of direction-dependent mock data cubes,
which in turn requires the use of the peeling-off technique to obtain adequate
sample statistics. With peeling-off we lose information about the origin of 
individual photons however.}.
The observational profile is instead median stacked, with the motivation being
to prevent faint undetected companions from enhancing the signal in the profile outskirts.
In our simulation, we resolve $\approx 100$ satellites per projected annular bin of width
$0.5 \, \mathrm{arcsec}$, of which there are generally $\approx 20$ satellites that contain
stars \cite[see, e.g., figure 2 in][for a visual impression]{Mitchell18}. When looking at radial 
profiles (not in projection) of escaping \lya luminosity, we therefore find that
while median (as opposed to mean) stacking does reduce the contribution from satellites 
in the average profile (by effectively excluding the brightest satellites), the reduction is modest 
(roughly a factor two in the number of the escaping photons that escape from satellites, 
and a $25 \%$ reduction in the number of escaping photons that were originally emitted
within satellites before scattering). The contribution from diffuse inflowing gas
is insensitive to mean versus median stacking, and is therefore slightly up-weighted
in relative importance for a median stack (compared to the mean stack shown in Fig.~\ref{SB_profile}).

A more realistic comparison to the observed profile is shown in Appendix~\ref{ap_11_comp},
using mock data cubes orientated along specific lines of sight (rather than angle-averaging), 
median stacking (matching the observed procedure), and an explicit flux selection
(rather than simply averaging all simulations outputs over $3<z<4$).
The more realistically produced profile almost exactly resembles the simulation profile shown
in Fig.~\ref{SB_profile}.

\subsection{\lya spectral morphology}
\label{spectral_morphology_sec}

\begin{figure}
\includegraphics[width=20pc]{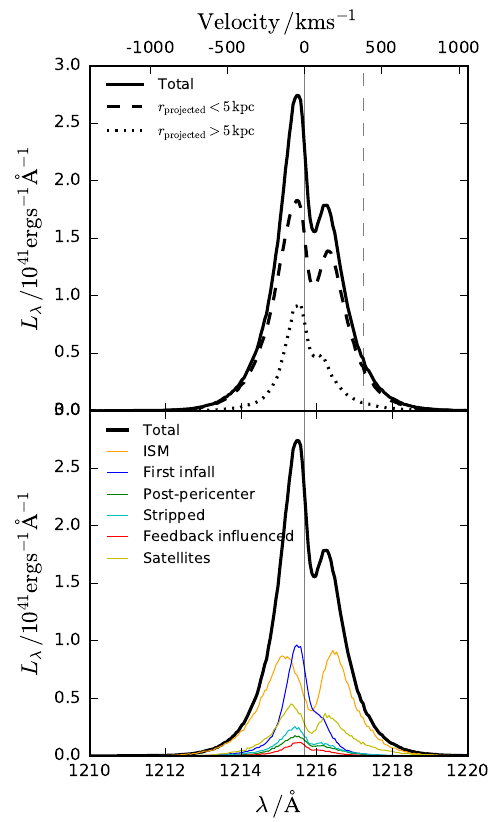}
\caption{Stacked ($3<z<4$), angle-averaged spectra of escaping \lya photons.
The top panel shows the relative spectral morphology of the inner versus outer region, split at $5 \, \mathrm{kpc}$ in projection.
The bottom panel shows the contribution of different components, splitting photons by which galaxy/CGM component each photon last interacted 
with before escaping, as indicated by the legend.
The dashed grey vertical line shown in the top panel indicates the peak shift of the \lya line predicted by the empirical relation from \protect \cite{Verhamme18}, given the full width at half maximum of our simulated spectrum for $r_{\mathrm{projected}} < 5 \, \mathrm{kpc}$.
Contrary to observed LAEs at $z \sim 3$ (for which IGM attenuation is expected to be modest), our simulated
spectrum exhibits flux bluewards of the systemic velocity that is comparable to the flux redwards of systemic.
As a consequence, the red peak of the simulated (inner region) spectrum is insufficiently redshifted relative to systemic, 
given the FWHM and the \protect \cite{Verhamme18} relation. 
}
\label{spectra}
\end{figure}

Fig.~\ref{spectra} shows the angle-averaged \lya spectrum of escaping photons for our simulation, mean stacked over $3<z<4$.
Since we are considering angle-averaged spectra, each \lya photon is adjusted in frequency to account for the peculiar
velocity of the halo within the cosmological box (otherwise the stacked spectrum would be artificially broadened).
The top panel of Fig.~\ref{spectra} shows the dichotomy between the spectrum of the spatial centre in projection
($r_{\mathrm{projected}} < 5 \, \mathrm{kpc}$),
and the spectrum of the spatially extended component. Most of the flux escapes from the central component, exhibiting
a double peaked spectral morphology, with a dominant blue peak slightly bluewards of the systemic velocity, 
and a more offset and a subdominant red peak. In contrast, the spatially extended component is asymmetric, single peaked, and peaking
slightly bluewards of systemic.

The lower panel of Fig.~\ref{spectra} shows the relative contribution of different Lagrangian components to
the overall profile (divided on the basis of which component each escaping photon last interacts with before
escaping the system). The spectral morphology of the flux tracing the central ISM (orange line) is the broadest
component, reflecting presumably the higher neutral HI optical depth of the dense ISM, thus requiring
larger frequency excursions before escape (though there is still significant flux escaping at the
systemic velocity, implying the existence of copious low-column density sight lines). The central ISM component is 
double-peaked, and near-symmetric around
systemic. \lya escaping from satellites (khaki line) is also double-peaked, but is less
symmetric. Satellites are responsible for the slight red ``bump'' feature that is apparent
for the spatially extended ($r_{\mathrm{projected}} > 5 \, \mathrm{kpc}$) component in the top panel. 
All of the other CGM components in the lower
panel are slightly blue-shifted and single peaked, explaining the overall elevation of blue
relative to red flux in the total spectrum shown in the top panel.

For the CGM-tracing components, an overall blueshift is expected for an inflowing medium 
\cite[e.g.][]{Dijkstra06,Verhamme06}, albeit with a fairly modest kinematic
shift due to the smaller expected absolute radial velocities of inflows compared to outflows. Typical radial inflow
velocities in our simulation are of order $50$ to $100 \, \mathrm{kms^{-1}}$ (see Fig.~\ref{phase_profiles}).
It is notable that the component tracing the gas affected by stellar feedback (red line) is
not clearly redshifted, given that feedback-driven outflows are commonly invoked to explain the 
systematically redshifted spectral morphology of observed LAEs \cite[e.g.][]{Ahn04, Verhamme06}.
Note that our definition of the ``feedback-influenced'' component includes gas that was 
affected by stellar feedback at any time in the past however, and is therefore comprised of
a mix of both inflowing and outflowing gas. Nonetheless, the lack of a strong redshifted
component reflects that fast-moving outflows in our simulation are generally fully ionized (see,
e.g., Fig.~\ref{fate_neut_outflows} and Fig.~\ref{phase_profiles}),
and are therefore transparent to \lya radiation.

Our simulated spectrum does not resemble the typical \lya spectrum of observed high-redshift galaxies. 
Observed \lya spectra (usually measured in an aperture that is comparable in size to the central ISM)
are nearly always observed with a red peak that is substantially redshifted
relative to systemic, are asymmetric with an extended tail towards the red, occasionally show a secondary blue peak,
and often show a strong deficit of photons at the systemic redshift if the secondary peak is present 
\cite[e.g.][]{Shapley03,Tapken07,Kulas12,Trainor15}.
There are observed counter-examples: for example a LAE with a double peak, a dominant blue peak and significant
emission at the systemic velocity was reported recently by \cite{Erb19}, and objects similar to our stacked spectrum
(i.e. are more symmetric around the systemic velocity) are occasionally seen 
\cite[e.g.][]{Tapken07,Erb16}, but are not representative.
Recently, \cite{Hayes20} presented mean and median stacks of LAEs from MUSE observations (albeit for LAEs that are generally
brighter than the object studied here), demonstrating that any flux bluewards of systemic is completely
subdominant to a dominant redshifted component. Furthermore, using observed \lya absorption statistics,
they find that the intervening IGM has a modest $\approx 25 \%$ attenuation of emission bluewards
of the systemic velocity at $z \sim 3$, and furthermore demonstrate that this is consistent with a comparison
with low-z LAEs, for which IGM attenuation is thought to be negligible. Simulation-based 
studies have reported slightly stronger IGM attenuation at $z \sim 3$ however \cite[e.g.][]{Byrohl20}.

\cite{Verhamme18} show that a positive correlation exists between the \lya FWHM, and the velocity shift of
peak from the systemic redshift of the galaxy. Given the FWHM of our stacked spectrum, we plot the predicted
\lya peak velocity ($V_{\mathrm{peak}} \approx 400 \, \mathrm{km s^{-1}}$) as the dashed grey vertical line in the top
panel of Fig.~\ref{spectra}. 
While there is systematic scatter in the \cite{Verhamme18} empirical relation, this serves to underline
that even for the ISM-tracing component, we likely significantly underestimate the velocity shift of the red peak
in the spectrum.

Another consideration is the comparison of the \lya spectral morphology between the spatial center
and the spatially extended halo.
Recent spatially resolved observations of high-redshift \lya spectra
reveal significant diversity in \lya halo spectra (the halo can be both red or blue-shifted with
respect to the central region), but that the halo spectra are still significantly 
redshifted from the systemic velocity of the system \cite[][]{Swinbank07,Smit17,Claeyssens19,Leclercq20},
although \cite{Erb18} present a more complex case where a double-peaked morphology is
seen in the outer halo.
These observations indicate that the unrealistic spectral morphology of our simulated galaxy
is therefore not likely confined to a problem on scales within the central ISM alone, though it is still
conceivable that the ISM is where large velocity shifts are imparted if most of the extended
emission is scattered from the CGM after being emitted within the ISM.

While it is possible that our simulated galaxy is an outlier in some way relative to the broader population, the differences 
between (typical) observed spectra with our simulation more likely imply that we do not have sufficient
HI column densities, and that we are missing a strongly outflowing and volume-filling neutral gas component in our simulation.
We note that if we inspect viewing angle-specific spectra (using the peeling off technique)
instead of angle-averaged spectra, we do see a very significant diversity in the spectral
morphologies (even if viewed from different angles at a single redshift), including
rare cases where there is a dominant redshifted component. We do not however find examples of spectra with 
negligible escaping emission bluewards of the systemic velocity, reinforcing the tension 
between our stacked angle-averaged spectrum and observations.

\section{Discussion}
\label{discussion_section}

\subsection{\lya spectra from cosmological simulations}

Given that our simulation fails on average to produce the characteristic single peaked, redshifted from systemic, and 
asymmetric spectral profile characteristic of observed LAEs at $z \sim 3$, we here review results presented
from other cosmological simulations in the literature, to assess the generality of the problem with simulations.
We focus on analyses of simulations that attempt \lya radiative transfer within the ISM, and that account (with
varying degrees of sophistication) for ionizing UV radiation from local stellar sources.

\cite{Smith19} present an analysis of a single cosmological zoom-in simulation from the FIRE simulation suite
\cite[][]{Hopkins14} for redshifts $z>5$, with comparable resolution to the simulation presented here.
Their figure 11 shows that before IGM attenuation is applied, 
their spectra also show strong blue peaks, and also high flux levels at the systemic velocity between the two 
peaks, similar to the ISM-tracing component for our galaxy shown in Fig.~\ref{spectra}. As they focus
on higher redshifts, they are able to achieve reasonable looking single-peaked redshifted \lya spectra
after IGM attenuation is applied, as the IGM attenuates almost all of the flux at line centre and on
the blue side. IGM attenuation is not likely to be a sufficient explanation at lower redshifts however \cite[e.g.][]{Dijkstra07,Laursen11, Inoue14, Hayes20},
and it would be interesting to see analysis of the \lya spectra from the FIRE simulations presented
at lower redshifts.

\cite{Behrens19} analyse a single cosmological zoom-in simulation run with the \ramses code (without
radiation hydrodynamics). Again, as their analysis is presented at $z=7$ the IGM completely attenuates
the \lya spectrum both at, and bluewards of the systemic velocity. Before their IGM correction their
spectra appear dominated by emission from a blue peak. 
\cite{Laursen07}, \cite{Laursen09} and \cite{Laursen09b} present analyses of an older and lower resolution (gas particle mass $\sim 10^5 \, \mathrm{M_\odot}$)
cosmological zoom-in simulation of a set of galaxies at $z=3.6$, and find similar results to \cite{Smith19}, 
in that the escaping \lya spectrum is double-peaked, with (in most cases) a stronger blue peak, 
and with substantial flux emerging at the systemic velocity.

As a counter-example, while not a live cosmological simulation, \cite{Verhamme12} analysed the emerging \lya 
spectra from two idealised low-mass disk galaxies. When the ISM was allowed to cool radiatively below $10^4 \, \mathrm{K}$ 
(their ``G2'' simulation), the emerging \lya profile is double-peaked when viewed edge-on, but is
asymmetric when viewed face-on, and with a single peak that is located redwards of the systemic velocity,
as is typical in real observations. This is attributed to outflowing gas in the simulation, but
it is not clear how the outflows in their simulation compare to the full cosmological simulations
discussed previously.
A second counter-example is provided by the recent analysis of a comparatively low-resolution cosmological
simulation by \cite{Chung19}, who do find redshifted spectra. In this case however, they utilise a subgrid
feedback model that temporarily decouples ``wind'' particles from the hydrodynamical scheme,
and ejects these particles from the ISM at high velocity. In such a scheme, supernova-driven
outflows can remain in a neutral phase even while outflowing at many hundreds of $\mathrm{km s^{-1}}$,
presumably explaining the difference in the predicted spectral morphology with the other
cosmological simulations discussed.

It can be argued that cosmological simulations produce unrealistic \lya spectral morphologies
because they do not sufficiently resolve the internal structure of star-forming clouds within the ISM,
where many of the photons are produced (in particular the Stromgren radius and
turbulent structure will be under-resolved). Using idealised radiation hydrodynamics simulations at 
$\sim 0.25 \, \mathrm{pc}$ resolution (a factor $50$ higher than for the cosmological simulation studied
here), \cite{Kimm19} study the escape of \lya photons
from individual star-forming clouds, 
isolated from the effects of the surrounding diffuse ISM and CGM. They find that radiation 
feedback from young stars can drive cloud expansion and
so create red-peak dominated \lya spectra, though they also show that if star formation efficiencies are
high ($10 \%$) clouds are disrupted early, leading to luminosity-weighted, time integrated
\lya spectra that are narrow and not kinematically shifted. \cite{Kimm19} also find that 
they not succeed in reproducing the large velocity shifts of the red peak that are often 
observed, concluding that larger-scale radiation 
transfer effects in the ISM/CGM are required.


Putting all of these examples together, we conclude that we have yet to see clear evidence
that current state-of-the-art cosmological simulations are capable of producing
realistic \lya spectra, unless wind particles are explicitly decoupled from
the hydrodynamical scheme. This point has also been discussed by \cite{Gronke17},
who interpret this problem as evidence that simulations indeed are not resolving
an important population of tiny optically thick clumps, concluding that sub-resolution
models for unresolved clumping in the ISM/CGM should be implemented when post-processing cosmological
simulations with \lya radiative transfer.
The relative lack of spatial resolution in the CGM has in particular received
a great deal of attention recently \cite[][]{Hummels19,Peeples19,Suresh19,vandevoort19},
and may be particularly pertinent given that we find that most of the escaping \lya
radiation from our galaxy does not scatter from the CGM (orange line, top panel, Fig.~\ref{SB_profile}). 
Another aspect that has received attention is the potential for cosmic rays to play
an important role in regulating escaping \lya spectra \cite[][]{Gronke18}, as cosmic rays can be
more resistant to dissipative losses and so more smoothly distribute the injected
feedback energy and momentum (both spatially and temporally), leading to less impulsive
but more persistent galactic-scale outflows that are better able to entrain and
maintain neutral hydrogen clouds.

At the same time, the unrealistic spectral morphology of our simulated galaxy
needs to be reconciled against the realistic \lya surface brightness profile,
which closely reproduces stacked MUSE observations 
(Fig.~\ref{SB_profile} and Fig.~\ref{lya_sb_mock}). At face value, adding
copious amounts of outflowing neutral gas to the CGM of our simulation
could lead to an over-production of \lya surface brightness at larger
projected separations. We do however under-predict the surface
brightness at projected separations of $\sim 15 \, \mathrm{kpc}$
by about $50 \, \%$, leaving room to add outflowing neutral gas
at these scales to act as a scattering medium. Recently, \cite{Song20}
have demonstrated that it is possible to reproduce both the surface brightness
profiles and spectral morphology of MUSE-observed LAEs, using idealised
spherically-symmetric models that feature a continuous expanding medium extending
outwards to scales of $\sim 20 \mathrm{kpc}$ with outflowing velocities 
$\sim 200 \, \mathrm{kms^{-1}}$ (their table 3). In future work we 
plan to see if similar results can be obtained empirically by modifying our simulations
in post-processing such that they they contain trace amounts of neutral
hydrogen in diffuse outflowing winds.

\subsection{Producing realistic stellar distributions in simulations of high-redshift galaxies}
\label{mhalo_problem}

\begin{figure}
\includegraphics[width=20pc]{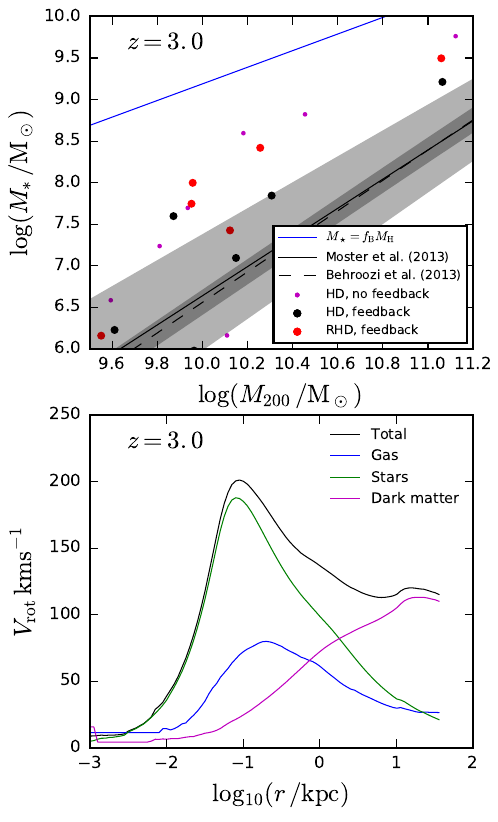}
\caption{Basic properties of our simulated galaxy.
\emph{\textbf{Top}:} stellar mass as a function of dark matter halo mass, plotted at $z=3$.
Red points show galaxies from the fiducial RHD simulation used in this study (which uses a realistic time delay distribution for supernova explosions).
Magenta points show galaxies from a hydrodynamics-only simulation, performed without stellar feedback, using the same initial conditions \protect \cite[][]{Mitchell18}.
Black points show galaxies from a corresponding hydrodynamics-only simulation that includes stellar feedback, using a fixed time delay of $10 \, \mathrm{Myr}$ for supernova explosions.
The main target galaxy is the most massive galaxy in each simulation with $M_{\mathrm{200}} \sim 10^{11.1} \, \mathrm{M_\odot}$
(other points are from neighbouring lower-mass galaxies within the zoom-in region).
Extrapolated constraints from abundance matching are shown by the grey lines bands, and are taken from \protect \cite{Behroozi13} and \protect \cite{Moster13}.
The blue diagonal line indicates the point where stellar mass would be equivalent to halo mass multiplied by the cosmic baryon fraction.
Our simulated galaxy is uncomfortably high in stellar mass given the halo mass, with any reduction in star formation from including RHD seemingly more than offset by using
a more realistic time delay distribution function for supernova explosions (less temporally clustered supernovae apparently result in higher radiative losses).
\emph{\textbf{Bottom}:} Rotation curve ($V_{\mathrm{rot}} \equiv G \, M(r) / r$) of the main galaxy from our fiducial RHD simulation at $z=3$, 
decomposed into the contribution from stars, gas and dark matter.
The rotation curve shows that the stellar distribution is highly centrally concentrated, indicative of likely radiative overcooling.}
\label{mstar_mhalo_vrot}
\end{figure}

An important caveats in our comparison to observed \lya surface brightness profiles 
is the question of whether we are simulating a LAE with the correct halo mass 
\cite[see for example][who predict that the in-situ \lya luminosity of the CGM strongly correlates with 
halo mass]{Rosdahl12}.
The top panel of Fig.~\ref{mstar_mhalo_vrot} shows the stellar mass of our simulated galaxy 
(red point at $M_{200} = 10^{11.1} \, \mathrm{M_\odot}$), compared 
to extrapolations of empirical constraints from \cite{Behroozi13} and \cite{Moster13}. 
With a stellar mass of $M_\star = 10^{9.5} \, \mathrm{M_\odot}$, our simulated
galaxy has an uncomfortably high stellar mass given its halo mass, sitting
an order of magnitude above the extrapolated best-fit empirical relations.

We also show the same galaxy but run without radiation hydrodynamics, both for the case of no stellar
feedback (magenta points) and with supernova feedback using a fixed time delay of $10 \, \mathrm{Myr}$ for
supernova explosions (black points). As a reminder, the fiducial RHD simulation used in this
study includes radiation feedback from local stellar sources, boosts the injected supernova momentum
to account for unresolved photo-heating \cite[][]{Geen15}, and uses a more realistic time
delay distribution function for supernova explosions. Excluding the boost associated with unresolved
photo-heating, both the RhD and non-RhD simulations with feedback inject four times the supernova 
energy expected for a Kroupa IMF (see Section~\ref{sec_sf_fb}).
While the RHD simulation forms about half the
mass in stars compared to the no-feedback case, it forms twice as many stars as the pure hydrodynamics
simulation with a fixed time delay. Evidently the expected reduction in stars formed due
to increased momentum input plus radiation feedback is more than offset by reducing
the temporal clustering of supernova explosions, which presumably increases the radiative cooling losses
\cite[somewhat contrary to expectations from recent work on idealised simulations, e.g., ][]{Gentry20,Keller20}.

In addition, the lower panel of Fig.~\ref{mstar_mhalo_vrot} shows the galaxy rotation curve (assuming
spherical symmetry), decomposed into contributions from stars, gas, and dark matter. The stellar component
is highly centrally concentrated, such that the resulting rotation curve is far from flat. While such
rotation curves cannot be currently ruled out for faint galaxies at $z>3$, we nonetheless interpret
this as an additional sign that the simulation likely suffers from overly high radiative losses relative to the energy
injected by feedback. We note that at this halo mass we are on the cusp of the regime where
feedback from a supermassive black hole may start to become relevant \cite[][]{Dubois15}, which may be an important
missing component needed to prevent the formation of such a compact stellar core (and could
in principle also act as a mechanism to accelerate dense gas to the velocities required to explain observed \lya spectral morphologies).

If it is the case that our galaxy has a stellar mass that is overly large
relative to its halo mass, the implication would be that the relative contribution of
photo-heating from local stellar sources to \lya emission could be over-estimated relative
to gravitional heating, and relative to heating from the wider UVB. This
is not guaranteed however, since more efficient stellar feedback (needed
to reduce the stellar mass) would also reduce the masses
of neutral atomic Hydrogen and dust within the ISM, feasibly increasing
\lya escape fractions for centrally emitted photons (which can then
scatter from the CGM).
Furthermore, since galaxy stellar mass scales approximately
as halo mass squared in the halo mass range of our simulated galaxy \cite[e.g.][]{Behroozi13,Moster13},
a simulation with more efficient feedback would only require a relatively modest change in 
halo mass to recover the same stellar mass (and therefore overall \lya brightness).
As such, we do not expect the predicted contribution of gravitional
heating to \lya production \cite[which is expected to scale linearly with halo mass,][]{Dijkstra09,Rosdahl12} 
to be strongly sensitive to the efficiency of feedback in our simulation.


\section{Summary}
\label{summary_sec}

In this study we set out to explore the origin of dense gas in the
circum-galactic medium (CGM) at high-redshift ($z \sim 3$), and the 
implications for spatially extended \lya emission from the CGM.
We have presented results taken from a radiation hydrodynamics simulation of a
high-redshift galaxy with a halo mass of $\sim 10^{11} \, \mathrm{M_\odot}$ at $z=3$.

We find that the spherically averaged radial mass profile of
circum-galactic neutral hydrogen in the simulation drops strongly with radius 
(Fig.~\ref{mp_profiles}), and that
this profile is shaped primarily by the compression of cosmologically
infalling gas pushing the typical hydrogen number density above
the threshold for efficient self-shielding from ionizing radiation 
(Fig.~\ref{fate_neut_outflows} and Fig.~\ref{cgm_profiles_lagrangian}). 
We also demonstrated that the
(subdominant) component of neutral outflowing circum-galactic hydrogen
is primarily comprised of gas that is still undergoing gravitational
infall, but has moved past first-pericentric passage, and will subsequently
fall back towards the ISM or settle into a approximate rotational equilibrium
within the inner CGM (Fig.~\ref{fate_neut_outflows} and Fig.~\ref{cgm_profiles_lagrangian}).

We then explore the implications of this scenario for the extended \lya
haloes recently detected around faint $z \sim 3$ \lya emitting galaxies
(LAEs) with the MUSE
instrument \cite[][]{Wisotzki16,Leclercq17}. We find that \lya emitted
in-situ within the CGM roughly (but not exactly) follows the radial 
profile of neutral hydrogen in the simulation (Fig.~\ref{cgm_profiles_lagrangian}),
and traces primarily dense and partially ionized infalling gas, 
spanning a range of densities and ionization conditions (Fig.~\ref{phase_profiles}). 
We find that angle-averaged emission from the CGM does
not vary significantly in time for our simulated galaxy, with variations
of a factor of a few at most (Fig.~\ref{cgm_origin}). This implies that high-redshift galaxies
may be generally surrounded by a diffuse \lya halo, irrespective
of the \lya escaping from the central region.

Combining in-situ emission from the CGM with emission emitted within
the central ISM and satellites, and accounting for the MUSE PSF, 
we find that scattered photons from the central ISM, in-situ emission
and emission from satellites contribute comparably to the spatially
extended signal (Fig.~\ref{SB_profile}). Scattering dominates
the inner $\mathrm{arcsec}$ of the \lya profile, and 
satellites provide much
of the very extended signal (beyond the spatial scales probed by observations
of individual LAEs, but which are accessibly via stacking).
Extended emission generally 
last traces inflowing gas before escaping the halo, or otherwise escapes 
from satellite galaxies without scattering from the CGM.
About $60 \%$ of the escaping photons are generated in recombinations,
with the remainder generated by collisional excitations (Fig.~\ref{rec_col_split}).
Compared to a stack of MUSE-detected LAEs from \cite{Wisotzki18},
the time and angle-averaged surface brightness profile from
our simulation reproduces the observations to at least $0.2 \, \mathrm{dex}$
over two orders of magnitude in surface brightness
(though admittedly the central $\mathrm{arcsec}$ of the profile
is severly affected by by PSF-broadening), out to beyond
the halo virial radius (Fig.~\ref{SB_profile}, but see also Appendix~\ref{ap_11_comp}).
 
Finally, we find that the average \lya spectrum produced by our galaxy
is very different from the typical spectrum of an observed LAE (Fig.~\ref{spectra}).
We find a complex spectral morphology with a dominant peak slightly bluewards
of systemic. The total spectrum is composed of a double-peaked spectral
component contributed by photons that escape from the ISM without scattering from
the CGM, and of a blue single peaked spectral component that traces inflowing
circum-galactic gas. This contrasts strongly with the spectral morphology
of typical high-redshift LAEs in observations; the spectra of observed LAEs 
at $z \sim 3$ are
usually significantly redshifted from the systemic velocity,
are asymmetric with a broad redward tail, and have minimal observed
flux bluewards of the systemic velocity (even accounting for IGM attenuation).

This problem seems to apply to other cosmological simulations
before IGM attenuation is applied \cite[e.g.][]{Behrens19,Smith19}, 
but since in our case we analyse
emission at redshifts $3<z<4$, IGM attenuation is unlikely to be a sufficient
explanation. We speculate that this is indeed a general problem
with current simulations 
\cite[see Section~\ref{discussion_section} as well as the discussion in][]{Gronke17},
and that the problems could be related to insufficient resolution in the CGM,
and/or because cosmological simulations do not (usually) include cosmic
ray creation and transport.

We conclude that observations of spatially resolved \lya emission
from high-redshift galaxies are providing highly challenging
constraints for state-of-the-art cosmological simulations.
These constraints are particularly timely given the current
debate regarding the sensitivity of the phase distribution of
the simulated CGM to numerical resolution, and the question
of whether cosmic rays play a significant role in shaping
the ISM and CGM of galaxies. With the upcoming launch of JWST
set to constrain whether high-redshift LAEs are surrounded by extended
haloes of Balmer-line emission, it will soon be possible to
establish whether in-situ emission from the CGM provides
a significant contribution to the extended signal.
If confirmed, and given that the extended emission is generally observed to be red-shifted
from systemic \cite[][]{Swinbank07,Smit17,Claeyssens19,Leclercq20}, the implication would then be that the red-shifted 
spectral morphology of high-redshift \lya lines is indeed being shaped by 
a fast-moving and neutral phase of CGM-scale (as opposed to only
ISM-scale) galactic outflows
with a high covering fraction, a component that is seemingly 
missing from current state-of-the-art cosmological simulations.

\section*{Acknowledgements}

We would like to thank Taysun Kimm and Anne Verhamme for providing comments and suggestions
that improved the quality of this manuscript.
We are grateful to the LABEX Lyon Institute of Origins (ANR-10-LABX-0066) of the Université de Lyon for its financial support within the program ``Investissements d'Avenir'' (ANR-11-IDEX-0007) of the French government operated by the National Research Agency (ANR).
This work was supported by Vici grant 639.043.409 from
the Netherlands Organisation for Scientific Research (NWO).
Simulations were run at the Common Computing Facility (CCF) of LABEX LIO. 
CC acknowledges support from the Institut Lagrange de Paris and from
the European Union Horizon 2020 research and innovation programme under
grant agreement No. 818085 GMGalaxies.

\section*{Data availability}

The data underlying this article will be shared on reasonable request to the corresponding author.

\bibliographystyle{mn2e}
\bibliography{bibliography}

\appendix
\input{ap_11_comp}

\label{lastpage}
\end{document}

%% file: ap_11_comp.tex
\section{Comparing to stacked \lya surface brightness profiles from MUSE observations}
\label{ap_11_comp}

\begin{figure}
\includegraphics[width=20pc]{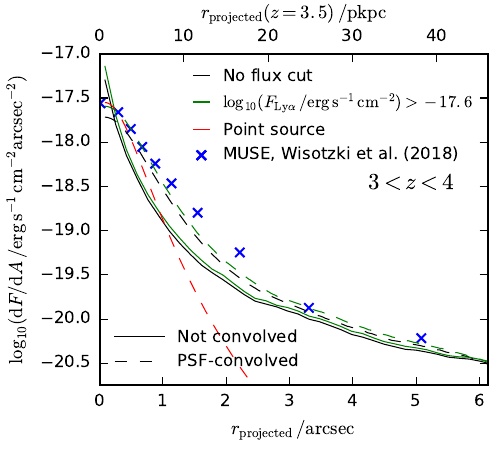}
\caption{A comparison of the predicted \lya surface brightness profile from our simulation
(for $3<z<4$) compared to the stack of LAEs produced using deep MUSE data by \protect \cite{Wisotzki18}.
In this case we more closely reproduce the observational selection and methodology by median stacking (after azimuthal averaging),
by using mock data cubes orientated along random lines of sight (rather than angle averaging) 
and by applying a flux cut to mimic the sensitivity of MUSE in detecting LAEs.
Solid (dashed) lines show profiles without (with) convolution with the appropriate MUSE point spread function (PSF).
The red dashed line shows the convolved profile of a point source for comparison, scaled to match the total luminosity
of the simulation stack when no flux cut is applied.
We choose a flux cut of $\log_{10}(F_{\mathrm{\lya}} \, / \mathrm{erg s^{-1} cm^{-2}}) > -17.6$
in order to match the peak surface brightness of the observed stack, which is comparable
to the approximate flux limit for which the MUSE observations are complete \protect \cite[see, e.g., figure 1 in][]{Leclercq17}.
With this flux cut, we reproduce the observed profile to within at least $0.2 \, \mathrm{dex}$
over the plotted range.
}
\label{lya_sb_mock}
\end{figure}

In Section~\ref{MUSE_comp_sec}, we compared a mean-stacked and angle-averaged surface brightness
profile of our simulated galaxy with a stack of observed LAEs from \cite{Wisotzki18}.
In Fig.~\ref{lya_sb_mock}, we perform a more realistic comparison with the observations by median stacking
(after first azimuthal averaging), by using mock data cubes orientated along $12$ random lines of 
sight for each simulation output (rather than angle averaging), and by applying a flux cut 
to mimic the sensitivity of MUSE in detecting LAEs.
We choose a flux cut of $\log_{10}(F_{\mathrm{\lya}} \, / \mathrm{erg s^{-1} cm^{-2}}) > -17.6$
in order to match the peak surface brightness of the observed stack, which is comparable
to the approximate flux limit for which the MUSE observations are complete \protect \cite[see, e.g., figure 1 in][]{Leclercq17}.
With this flux cut, we reproduce the observed profile to within at least $0.2 \, \mathrm{dex}$
over the plotted range. 

The mock data cubes used here make use of the peeling off technique in order to obtain
sufficient signal along specific lines of sight (which in turn is needed to apply an observational
flux cut). By using peeling off, we lose the information about the origin of each individual photon,
which is why we do not use the mock cubes for the results presented in \protect Sec~\ref{flows_sec}.
With the adopted flux cut, the simulated surface brightness profile in any case very closely resembles
the mean, angle-averaged stack shown in \protect Fig.~\ref{SB_profile}.

%% file: tracing_lya.bbl
\begin{thebibliography}{128}
\expandafter\ifx\csname natexlab\endcsname\relax\def\natexlab#1{#1}\fi

\bibitem[{{Ahn}(2004)}]{Ahn04}
{Ahn} S.-H., 2004, \apjl, 601, L25

\bibitem[{{Aubert} {et~al}\mbox{.}(2004){Aubert}, {Pichon}, \&
  {Colombi}}]{Aubert04}
{Aubert} D., {Pichon} C., {Colombi} S., 2004, \mnras, 352, 376

\bibitem[{{Bacon} {et~al}\mbox{.}(2017){Bacon}, {Conseil}, {Mary},
  {Brinchmann}, {Shepherd}, {Akhlaghi}, {Weilbacher}, {Piqueras}, {Wisotzki},
  {Lagattuta}, {Epinat}, {Guerou}, {Inami}, {Cantalupo}, {Courbot}, {Contini},
  {Richard}, {Maseda}, {Bouwens}, {Bouch{\'e}}, {Kollatschny}, {Schaye},
  {Marino}, {Pello}, {Herenz}, {Guiderdoni}, \& {Carollo}}]{Bacon17}
{Bacon} R. {et~al.}, 2017, \aap, 608, A1

\bibitem[{{Barnes} {et~al}\mbox{.}(2011){Barnes}, {Haehnelt}, {Tescari}, \&
  {Viel}}]{Barnes11}
{Barnes} L.~A., {Haehnelt} M.~G., {Tescari} E., {Viel} M., 2011, \mnras, 416,
  1723

\bibitem[{{Behrens} {et~al}\mbox{.}(2019){Behrens}, {Pallottini}, {Ferrara},
  {Gallerani}, \& {Vallini}}]{Behrens19}
{Behrens} C., {Pallottini} A., {Ferrara} A., {Gallerani} S., {Vallini} L.,
  2019, \mnras, 486, 2197

\bibitem[{{Behroozi} {et~al}\mbox{.}(2013){Behroozi}, {Wechsler}, \&
  {Conroy}}]{Behroozi13}
{Behroozi} P.~S., {Wechsler} R.~H., {Conroy} C., 2013, \apj, 770, 57

\bibitem[{{Binney} \& {Tremaine}(2008)}]{Binney08}
{Binney} J., {Tremaine} S., 2008, {Galactic Dynamics: Second Edition}.
  Princeton University Press

\bibitem[{{Blondin} {et~al}\mbox{.}(1998){Blondin}, {Wright}, {Borkowski}, \&
  {Reynolds}}]{Blondin98}
{Blondin} J.~M., {Wright} E.~B., {Borkowski} K.~J., {Reynolds} S.~P., 1998,
  \apj, 500, 342

\bibitem[{{Byrohl} \& {Gronke}(2020)}]{Byrohl20}
{Byrohl} C., {Gronke} M., 2020, \aap, 642, L16

\bibitem[{{Cadiou} {et~al}\mbox{.}(2019){Cadiou}, {Dubois}, \&
  {Pichon}}]{Cadiou19}
{Cadiou} C., {Dubois} Y., {Pichon} C., 2019, \aap, 621, A96

\bibitem[{{Cantalupo} {et~al}\mbox{.}(2014){Cantalupo}, {Arrigoni-Battaia},
  {Prochaska}, {Hennawi}, \& {Madau}}]{Cantalupo14}
{Cantalupo} S., {Arrigoni-Battaia} F., {Prochaska} J.~X., {Hennawi} J.~F.,
  {Madau} P., 2014, \nat, 506, 63

\bibitem[{{Cantalupo} {et~al}\mbox{.}(2008){Cantalupo}, {Porciani}, \&
  {Lilly}}]{Cantalupo08}
{Cantalupo} S., {Porciani} C., {Lilly} S.~J., 2008, \apj, 672, 48

\bibitem[{{Cantalupo} {et~al}\mbox{.}(2005){Cantalupo}, {Porciani}, {Lilly}, \&
  {Miniati}}]{Cantalupo05}
{Cantalupo} S., {Porciani} C., {Lilly} S.~J., {Miniati} F., 2005, \apj, 628, 61

\bibitem[{{Chisholm} {et~al}\mbox{.}(2016){Chisholm}, {Tremonti Christy},
  {Leitherer}, \& {Chen}}]{Chisholm16}
{Chisholm} J., {Tremonti Christy} A., {Leitherer} C., {Chen} Y., 2016, \mnras,
  463, 541

\bibitem[{{Chung} {et~al}\mbox{.}(2019){Chung}, {Dijkstra}, {Ciardi},
  {Kakiichi}, \& {Naab}}]{Chung19}
{Chung} A.~S., {Dijkstra} M., {Ciardi} B., {Kakiichi} K., {Naab} T., 2019,
  \mnras, 484, 2420

\bibitem[{{Claeyssens} {et~al}\mbox{.}(2019){Claeyssens}, {Richard}, {Blaizot},
  {Garel}, {Leclercq}, {Patr{\'\i}cio}, {Verhamme}, {Wisotzki}, {Bacon},
  {Carton}, {Cl{\'e}ment}, {Herenz}, {Marino}, {Muzahid}, {Saust}, \&
  {Schaye}}]{Claeyssens19}
{Claeyssens} A. {et~al.}, 2019, \mnras, 489, 5022

\bibitem[{{Dekel} \& {Birnboim}(2006)}]{Dekel06}
{Dekel} A., {Birnboim} Y., 2006, \mnras, 368, 2

\bibitem[{{D{\'\i}az} {et~al}\mbox{.}(2020){D{\'\i}az}, {Ryan-Weber}, {Karman},
  {Caputi}, {Salvadori}, {Crighton}, {Ouchi}, \& {Vanzella}}]{Diaz20}
{D{\'\i}az} C.~G., {Ryan-Weber} E.~V., {Karman} W., {Caputi} K.~I., {Salvadori}
  S., {Crighton} N.~H., {Ouchi} M., {Vanzella} E., 2020, \mnras

\bibitem[{{Dijkstra}(2017)}]{Dijkstra17}
{Dijkstra} M., 2017, arXiv e-prints, arXiv:1704.03416

\bibitem[{{Dijkstra} {et~al}\mbox{.}(2006){Dijkstra}, {Haiman}, \&
  {Spaans}}]{Dijkstra06}
{Dijkstra} M., {Haiman} Z., {Spaans} M., 2006, \apj, 649, 14

\bibitem[{{Dijkstra} \& {Kramer}(2012)}]{Dijkstra12}
{Dijkstra} M., {Kramer} R., 2012, \mnras, 424, 1672

\bibitem[{{Dijkstra} {et~al}\mbox{.}(2007){Dijkstra}, {Lidz}, \&
  {Wyithe}}]{Dijkstra07}
{Dijkstra} M., {Lidz} A., {Wyithe} J. S.~B., 2007, \mnras, 377, 1175

\bibitem[{{Dijkstra} \& {Loeb}(2009)}]{Dijkstra09}
{Dijkstra} M., {Loeb} A., 2009, \mnras, 400, 1109

\bibitem[{{Dubois} {et~al}\mbox{.}(2015){Dubois}, {Volonteri}, {Silk},
  {Devriendt}, {Slyz}, \& {Teyssier}}]{Dubois15}
{Dubois} Y., {Volonteri} M., {Silk} J., {Devriendt} J., {Slyz} A., {Teyssier}
  R., 2015, \mnras, 452, 1502

\bibitem[{{Eldridge} {et~al}\mbox{.}(2008){Eldridge}, {Izzard}, \&
  {Tout}}]{Elridge08}
{Eldridge} J.~J., {Izzard} R.~G., {Tout} C.~A., 2008, \mnras, 384, 1109

\bibitem[{{Eldridge} \& {Stanway}(2016)}]{Elridge16}
{Eldridge} J.~J., {Stanway} E.~R., 2016, \mnras, 462, 3302

\bibitem[{{Erb} {et~al}\mbox{.}(2019){Erb}, {Berg}, {Auger}, {Kaplan},
  {Brammer}, \& {Pettini}}]{Erb19}
{Erb} D.~K., {Berg} D.~A., {Auger} M.~W., {Kaplan} D.~L., {Brammer} G.,
  {Pettini} M., 2019, \apj, 884, 7

\bibitem[{{Erb} {et~al}\mbox{.}(2016){Erb}, {Pettini}, {Steidel}, {Strom},
  {Rudie}, {Trainor}, {Shapley}, \& {Reddy}}]{Erb16}
{Erb} D.~K., {Pettini} M., {Steidel} C.~C., {Strom} A.~L., {Rudie} G.~C.,
  {Trainor} R.~F., {Shapley} A.~E., {Reddy} N.~A., 2016, \apj, 830, 52

\bibitem[{{Erb} {et~al}\mbox{.}(2018){Erb}, {Steidel}, \& {Chen}}]{Erb18}
{Erb} D.~K., {Steidel} C.~C., {Chen} Y., 2018, \apjl, 862, L10

\bibitem[{{Fardal} {et~al}\mbox{.}(2001){Fardal}, {Katz}, {Gardner},
  {Hernquist}, {Weinberg}, \& {Dav{\'e}}}]{Fardal01}
{Fardal} M.~A., {Katz} N., {Gardner} J.~P., {Hernquist} L., {Weinberg} D.~H.,
  {Dav{\'e}} R., 2001, \apj, 562, 605

\bibitem[{{Faucher-Gigu{\`e}re} {et~al}\mbox{.}(2010){Faucher-Gigu{\`e}re},
  {Kere{\v{s}}}, {Dijkstra}, {Hernquist}, \& {Zaldarriaga}}]{FaucherGiguere10}
{Faucher-Gigu{\`e}re} C.-A., {Kere{\v{s}}} D., {Dijkstra} M., {Hernquist} L.,
  {Zaldarriaga} M., 2010, \apj, 725, 633

\bibitem[{{Faucher-Gigu{\`e}re} {et~al}\mbox{.}(2009){Faucher-Gigu{\`e}re},
  {Lidz}, {Zaldarriaga}, \& {Hernquist}}]{FaucherGiguere09}
{Faucher-Gigu{\`e}re} C.-A., {Lidz} A., {Zaldarriaga} M., {Hernquist} L., 2009,
  \apj, 703, 1416

\bibitem[{{Federrath} \& {Klessen}(2012)}]{Federrath12}
{Federrath} C., {Klessen} R.~S., 2012, \apj, 761, 156

\bibitem[{{Ferland} {et~al}\mbox{.}(1998){Ferland}, {Korista}, {Verner},
  {Ferguson}, {Kingdon}, \& {Verner}}]{Ferland98}
{Ferland} G.~J., {Korista} K.~T., {Verner} D.~A., {Ferguson} J.~W., {Kingdon}
  J.~B., {Verner} E.~M., 1998, \pasp, 110, 761

\bibitem[{{Fujita} {et~al}\mbox{.}(2009){Fujita}, {Martin}, {Mac Low}, {New},
  \& {Weaver}}]{Fujita09}
{Fujita} A., {Martin} C.~L., {Mac Low} M.-M., {New} K. C.~B., {Weaver} R.,
  2009, \apj, 698, 693

\bibitem[{{Furlanetto} {et~al}\mbox{.}(2005){Furlanetto}, {Schaye}, {Springel},
  \& {Hernquist}}]{Furlanetto05}
{Furlanetto} S.~R., {Schaye} J., {Springel} V., {Hernquist} L., 2005, \apj,
  622, 7

\bibitem[{{Gallego} {et~al}\mbox{.}(2018){Gallego}, {Cantalupo}, {Lilly},
  {Marino}, {Pezzulli}, {Schaye}, {Wisotzki}, {Bacon}, {Inami}, {Akhlaghi},
  {Tacchella}, {Richard}, {Bouche}, {Steinmetz}, \& {Carollo}}]{Gallego18}
{Gallego} S.~G. {et~al.}, 2018, \mnras, 475, 3854

\bibitem[{{Geen} {et~al}\mbox{.}(2015){Geen}, {Rosdahl}, {Blaizot},
  {Devriendt}, \& {Slyz}}]{Geen15}
{Geen} S., {Rosdahl} J., {Blaizot} J., {Devriendt} J., {Slyz} A., 2015, \mnras,
  448, 3248

\bibitem[{{Genel} {et~al}\mbox{.}(2013){Genel}, {Vogelsberger}, {Nelson},
  {Sijacki}, {Springel}, \& {Hernquist}}]{Genel13}
{Genel} S., {Vogelsberger} M., {Nelson} D., {Sijacki} D., {Springel} V.,
  {Hernquist} L., 2013, \mnras, 435, 1426

\bibitem[{{Gentry} {et~al}\mbox{.}(2020){Gentry}, {Madau}, \&
  {Krumholz}}]{Gentry20}
{Gentry} E.~S., {Madau} P., {Krumholz} M.~R., 2020, \mnras, 492, 1243

\bibitem[{{Goerdt} {et~al}\mbox{.}(2010){Goerdt}, {Dekel}, {Sternberg},
  {Ceverino}, {Teyssier}, \& {Primack}}]{Goerdt10}
{Goerdt} T., {Dekel} A., {Sternberg} A., {Ceverino} D., {Teyssier} R.,
  {Primack} J.~R., 2010, \mnras, 407, 613

\bibitem[{{Gronke}(2017)}]{Gronke17b}
{Gronke} M., 2017, \aap, 608, A139

\bibitem[{{Gronke} {et~al}\mbox{.}(2017){Gronke}, {Dijkstra}, {McCourt}, \&
  {Peng Oh}}]{Gronke17}
{Gronke} M., {Dijkstra} M., {McCourt} M., {Peng Oh} S., 2017, \aap, 607, A71

\bibitem[{{Gronke} {et~al}\mbox{.}(2018){Gronke}, {Girichidis}, {Naab}, \&
  {Walch}}]{Gronke18}
{Gronke} M., {Girichidis} P., {Naab} T., {Walch} S., 2018, \apjl, 862, L7

\bibitem[{{Guillet} \& {Teyssier}(2011)}]{Guillet11}
{Guillet} T., {Teyssier} R., 2011, Journal of Computational Physics, 230, 4756

\bibitem[{{Haardt} \& {Madau}(1996)}]{Haardt96}
{Haardt} F., {Madau} P., 1996, \apj, 461, 20

\bibitem[{{Hahn} \& {Abel}(2011)}]{Hahn11}
{Hahn} O., {Abel} T., 2011, \mnras, 415, 2101

\bibitem[{{Haiman} \& {Rees}(2001)}]{Haiman01}
{Haiman} Z., {Rees} M.~J., 2001, \apj, 556, 87

\bibitem[{{Hashimoto} {et~al}\mbox{.}(2017){Hashimoto}, {Garel}, {Guiderdoni},
  {Drake}, {Bacon}, {Blaizot}, {Richard}, {Leclercq}, {Inami}, {Verhamme},
  {Bouwens}, {Brinchmann}, {Cantalupo}, {Carollo}, {Caruana}, {Herenz},
  {Kerutt}, {Marino}, {Mitchell}, \& {Schaye}}]{Hashimoto17}
{Hashimoto} T. {et~al.}, 2017, \aap, 608, A10

\bibitem[{{Hashimoto} {et~al}\mbox{.}(2015){Hashimoto}, {Verhamme}, {Ouchi},
  {Shimasaku}, {Schaerer}, {Nakajima}, {Shibuya}, {Rauch}, {Ono}, \&
  {Goto}}]{Hashimoto15}
{Hashimoto} T. {et~al.}, 2015, \apj, 812, 157

\bibitem[{{Hayes} {et~al}\mbox{.}(2013){Hayes}, {{\"O}stlin}, {Schaerer},
  {Verhamme}, {Mas-Hesse}, {Adamo}, {Atek}, {Cannon}, {Duval}, {Guaita},
  {Herenz}, {Kunth}, {Laursen}, {Melinder}, {Orlitov{\'a}},
  {Ot{\'\i}-Floranes}, \& {Sandberg}}]{Hayes13}
{Hayes} M. {et~al.}, 2013, \apjl, 765, L27

\bibitem[{{Hayes} {et~al}\mbox{.}(2020){Hayes}, {Runnholm}, {Gronke}, \&
  {Scarlata}}]{Hayes20}
{Hayes} M.~J., {Runnholm} A., {Gronke} M., {Scarlata} C., 2020, arXiv e-prints,
  arXiv:2006.03232

\bibitem[{{Heckman} {et~al}\mbox{.}(1991){Heckman}, {Lehnert}, {Miley}, \& {van
  Breugel}}]{Heckman91}
{Heckman} T.~M., {Lehnert} M.~D., {Miley} G.~K., {van Breugel} W., 1991, \apj,
  381, 373

\bibitem[{{Heckman} {et~al}\mbox{.}(2000){Heckman}, {Lehnert}, {Strickland}, \&
  {Armus}}]{Heckman00}
{Heckman} T.~M., {Lehnert} M.~D., {Strickland} D.~K., {Armus} L., 2000, \apjs,
  129, 493

\bibitem[{{Hopkins} {et~al}\mbox{.}(2014){Hopkins}, {Kere{\v s}}, {O{\~n}orbe},
  {Faucher-Gigu{\`e}re}, {Quataert}, {Murray}, \& {Bullock}}]{Hopkins14}
{Hopkins} P.~F., {Kere{\v s}} D., {O{\~n}orbe} J., {Faucher-Gigu{\`e}re} C.-A.,
  {Quataert} E., {Murray} N., {Bullock} J.~S., 2014, \mnras, 445, 581

\bibitem[{{Hopkins} {et~al}\mbox{.}(2018){Hopkins}, {Wetzel}, {Kere{\v{s}}},
  {Faucher-Gigu{\`e}re}, {Quataert}, {Boylan-Kolchin}, {Murray}, {Hayward},
  {Garrison-Kimmel}, {Hummels}, {Feldmann}, {Torrey}, {Ma},
  {Angl{\'e}s-Alc{\'a}zar}, {Su}, {Orr}, {Schmitz}, {Escala}, {Sanderson},
  {Grudi{\'c}}, {Hafen}, {Kim}, {Fitts}, {Bullock}, {Wheeler}, {Chan},
  {Elbert}, \& {Narayanan}}]{Hopkins18}
{Hopkins} P.~F. {et~al.}, 2018, \mnras, 480, 800

\bibitem[{{Hui} \& {Gnedin}(1997)}]{Hui97}
{Hui} L., {Gnedin} N.~Y., 1997, \mnras, 292, 27

\bibitem[{{Hummels} {et~al}\mbox{.}(2019){Hummels}, {Smith}, {Hopkins},
  {O'Shea}, {Silvia}, {Werk}, {Lehner}, {Wise}, {Collins}, \&
  {Butsky}}]{Hummels19}
{Hummels} C.~B. {et~al.}, 2019, \apj, 882, 156

\bibitem[{{Inami} {et~al}\mbox{.}(2017){Inami}, {Bacon}, {Brinchmann},
  {Richard}, {Contini}, {Conseil}, {Hamer}, {Akhlaghi}, {Bouch{\'e}},
  {Cl{\'e}ment}, {Desprez}, {Drake}, {Hashimoto}, {Leclercq}, {Maseda},
  {Michel-Dansac}, {Paalvast}, {Tresse}, {Ventou}, {Kollatschny}, {Boogaard},
  {Finley}, {Marino}, {Schaye}, \& {Wisotzki}}]{Inami17}
{Inami} H. {et~al.}, 2017, \aap, 608, A2

\bibitem[{{Inoue} {et~al}\mbox{.}(2014){Inoue}, {Shimizu}, {Iwata}, \&
  {Tanaka}}]{Inoue14}
{Inoue} A.~K., {Shimizu} I., {Iwata} I., {Tanaka} M., 2014, \mnras, 442, 1805

\bibitem[{{Keller} \& {Kruijssen}(2020)}]{Keller20}
{Keller} B.~W., {Kruijssen} J.~M.~D., 2020, arXiv e-prints, arXiv:2004.03608

\bibitem[{{Kere{\v s}} {et~al}\mbox{.}(2005){Kere{\v s}}, {Katz}, {Weinberg},
  \& {Dav{\'e}}}]{Keres05}
{Kere{\v s}} D., {Katz} N., {Weinberg} D.~H., {Dav{\'e}} R., 2005, \mnras, 363,
  2

\bibitem[{{Kimm} {et~al}\mbox{.}(2019){Kimm}, {Blaizot}, {Garel},
  {Michel-Dansac}, {Katz}, {Rosdahl}, {Verhamme}, \& {Haehnelt}}]{Kimm19}
{Kimm} T., {Blaizot} J., {Garel} T., {Michel-Dansac} L., {Katz} H., {Rosdahl}
  J., {Verhamme} A., {Haehnelt} M., 2019, \mnras, 486, 2215

\bibitem[{{Kimm} \& {Cen}(2014)}]{Kimm14}
{Kimm} T., {Cen} R., 2014, \apj, 788, 121

\bibitem[{{Kimm} {et~al}\mbox{.}(2015){Kimm}, {Cen}, {Devriendt}, {Dubois}, \&
  {Slyz}}]{Kimm15}
{Kimm} T., {Cen} R., {Devriendt} J., {Dubois} Y., {Slyz} A., 2015, \mnras, 451,
  2900

\bibitem[{{Kimm} {et~al}\mbox{.}(2017){Kimm}, {Katz}, {Haehnelt}, {Rosdahl},
  {Devriendt}, \& {Slyz}}]{Kimm17}
{Kimm} T., {Katz} H., {Haehnelt} M., {Rosdahl} J., {Devriendt} J., {Slyz} A.,
  2017, \mnras, 466, 4826

\bibitem[{{Kollmeier} {et~al}\mbox{.}(2010){Kollmeier}, {Zheng}, {Dav{\'e}},
  {Gould}, {Katz}, {Miralda-Escud{\'e}}, \& {Weinberg}}]{Kollmeier10}
{Kollmeier} J.~A., {Zheng} Z., {Dav{\'e}} R., {Gould} A., {Katz} N.,
  {Miralda-Escud{\'e}} J., {Weinberg} D.~H., 2010, \apj, 708, 1048

\bibitem[{{Kroupa}(2002)}]{Kroupa02}
{Kroupa} P., 2002, Science, 295, 82

\bibitem[{{Kulas} {et~al}\mbox{.}(2012){Kulas}, {Shapley}, {Kollmeier},
  {Zheng}, {Steidel}, \& {Hainline}}]{Kulas12}
{Kulas} K.~R., {Shapley} A.~E., {Kollmeier} J.~A., {Zheng} Z., {Steidel} C.~C.,
  {Hainline} K.~N., 2012, \apj, 745, 33

\bibitem[{{Lake} {et~al}\mbox{.}(2015){Lake}, {Zheng}, {Cen}, {Sadoun},
  {Momose}, \& {Ouchi}}]{Lake15}
{Lake} E., {Zheng} Z., {Cen} R., {Sadoun} R., {Momose} R., {Ouchi} M., 2015,
  \apj, 806, 46

\bibitem[{{Laursen} {et~al}\mbox{.}(2009{\natexlab{a}}){Laursen}, {Razoumov},
  \& {Sommer-Larsen}}]{Laursen09}
{Laursen} P., {Razoumov} A.~O., {Sommer-Larsen} J., 2009{\natexlab{a}}, \apj,
  696, 853

\bibitem[{{Laursen} \& {Sommer-Larsen}(2007)}]{Laursen07}
{Laursen} P., {Sommer-Larsen} J., 2007, \apjl, 657, L69

\bibitem[{{Laursen} {et~al}\mbox{.}(2009{\natexlab{b}}){Laursen},
  {Sommer-Larsen}, \& {Andersen}}]{Laursen09b}
{Laursen} P., {Sommer-Larsen} J., {Andersen} A.~C., 2009{\natexlab{b}}, \apj,
  704, 1640

\bibitem[{{Laursen} {et~al}\mbox{.}(2011){Laursen}, {Sommer-Larsen}, \&
  {Razoumov}}]{Laursen11}
{Laursen} P., {Sommer-Larsen} J., {Razoumov} A.~O., 2011, \apj, 728, 52

\bibitem[{{Leclercq} {et~al}\mbox{.}(2020){Leclercq}, {Bacon}, {Verhamme},
  {Garel}, {Blaizot}, {Brinchmann}, {Cantalupo}, {Claeyssens}, {Conseil},
  {Contini}, {Hashimoto}, {Herenz}, {Kusakabe}, {Marino}, {Maseda}, {Matthee},
  {Mitchell}, {Pezzulli}, {Richard}, {Schmidt}, \& {Wisotzki}}]{Leclercq20}
{Leclercq} F. {et~al.}, 2020, \aap, 635, A82

\bibitem[{{Leclercq} {et~al}\mbox{.}(2017){Leclercq}, {Bacon}, {Wisotzki},
  {Mitchell}, {Garel}, {Verhamme}, {Blaizot}, {Hashimoto}, {Herenz}, {Conseil},
  {Cantalupo}, {Inami}, {Contini}, {Richard}, {Maseda}, {Schaye}, {Marino},
  {Akhlaghi}, {Brinchmann}, \& {Carollo}}]{Leclercq17}
{Leclercq} F. {et~al.}, 2017, \aap, 608, A8

\bibitem[{{Levermore}(1984)}]{Levermore84}
{Levermore} C.~D., 1984, \jqsrt, 31, 149

\bibitem[{{Lofthouse} {et~al}\mbox{.}(2019){Lofthouse}, {Fumagalli}, {Fossati},
  {O'Meara}, {Murphy}, {Christensen}, {Prochaska}, {Cantalupo}, {Bielby},
  {Cooke}, {Lusso}, \& {Morris}}]{Lofthouse19}
{Lofthouse} E.~K. {et~al.}, 2019, \mnras, 2667

\bibitem[{{Mackenzie} {et~al}\mbox{.}(2019){Mackenzie}, {Fumagalli}, {Theuns},
  {Hatton}, {Garel}, {Cantalupo}, {Christensen}, {Fynbo}, {Kanekar},
  {M{\o}ller}, {O'Meara}, {Prochaska}, {Rafelski}, {Shanks}, \&
  {Trayford}}]{Mackenzie19}
{Mackenzie} R. {et~al.}, 2019, \mnras, 487, 5070

\bibitem[{{Mas-Ribas} \& {Dijkstra}(2016)}]{MasRibas16}
{Mas-Ribas} L., {Dijkstra} M., 2016, \apj, 822, 84

\bibitem[{{Mas-Ribas} {et~al}\mbox{.}(2017){Mas-Ribas}, {Dijkstra}, {Hennawi},
  {Trenti}, {Momose}, \& {Ouchi}}]{MasRibas17}
{Mas-Ribas} L., {Dijkstra} M., {Hennawi} J.~F., {Trenti} M., {Momose} R.,
  {Ouchi} M., 2017, \apj, 841, 19

\bibitem[{{Matsuda} {et~al}\mbox{.}(2012){Matsuda}, {Yamada}, {Hayashino},
  {Yamauchi}, {Nakamura}, {Morimoto}, {Ouchi}, {Ono}, {Umemura}, \&
  {Mori}}]{Matsuda12}
{Matsuda} Y. {et~al.}, 2012, \mnras, 425, 878

\bibitem[{{McCourt} {et~al}\mbox{.}(2018){McCourt}, {Oh}, {O'Leary}, \&
  {Madigan}}]{Mccourt18}
{McCourt} M., {Oh} S.~P., {O'Leary} R., {Madigan} A.-M., 2018, \mnras, 473,
  5407

\bibitem[{{Michel-Dansac} {et~al}\mbox{.}(2020){Michel-Dansac}, {Blaizot},
  {Garel}, {Verhamme}, {Kimm}, \& {Trebitsch}}]{MichelDansac20}
{Michel-Dansac} L., {Blaizot} J., {Garel} T., {Verhamme} A., {Kimm} T.,
  {Trebitsch} M., 2020, \aap, 635, A154

\bibitem[{{Mitchell} {et~al}\mbox{.}(2018){Mitchell}, {Blaizot}, {Devriendt},
  {Kimm}, {Michel-Dansac}, {Rosdahl}, \& {Slyz}}]{Mitchell18}
{Mitchell} P.~D., {Blaizot} J., {Devriendt} J., {Kimm} T., {Michel-Dansac} L.,
  {Rosdahl} J., {Slyz} A., 2018, \mnras, 474, 4279

\bibitem[{{Mori} {et~al}\mbox{.}(2004){Mori}, {Umemura}, \& {Ferrara}}]{Mori04}
{Mori} M., {Umemura} M., {Ferrara} A., 2004, \apjl, 613, L97

\bibitem[{{Moster} {et~al}\mbox{.}(2013){Moster}, {Naab}, \&
  {White}}]{Moster13}
{Moster} B.~P., {Naab} T., {White} S.~D.~M., 2013, \mnras, 428, 3121

\bibitem[{{Muzahid} {et~al}\mbox{.}(2020){Muzahid}, {Schaye}, {Marino},
  {Cantalupo}, {Brinchmann}, {Contini}, {Wendt}, {Wisotzki}, {Zabl},
  {Bouch{\'e}}, {Akhlaghi}, {Chen}, {Claeyssens}, {Johnson}, {Leclercq},
  {Maseda}, {Matthee}, {Richard}, {Urrutia}, \& {Verhamme}}]{Muzahid20}
{Muzahid} S. {et~al.}, 2020, \mnras, 496, 1013

\bibitem[{{Peeples} {et~al}\mbox{.}(2019){Peeples}, {Corlies}, {Tumlinson},
  {O'Shea}, {Lehner}, {O'Meara}, {Howk}, {Earl}, {Smith}, {Wise}, \&
  {Hummels}}]{Peeples19}
{Peeples} M.~S. {et~al.}, 2019, \apj, 873, 129

\bibitem[{{Planck Collaboration} {et~al}\mbox{.}(2014){Planck Collaboration},
  {Ade}, {Aghanim}, {Armitage-Caplan}, {Arnaud}, {Ashdown}, {Atrio-Barand ela},
  {Aumont}, {Baccigalupi}, {Banday}, {Barreiro}, {Bartlett}, {Battaner},
  {Benabed}, {Beno{\^\i}t}, {Benoit-L{\'e}vy}, {Bernard}, {Bersanelli},
  {Bielewicz}, {Bobin}, {Bock}, {Bonaldi}, {Bond}, {Borrill}, {Bouchet},
  {Bridges}, {Bucher}, {Burigana}, {Butler}, {Calabrese}, {Cappellini},
  {Cardoso}, {Catalano}, {Challinor}, {Chamballu}, {Chary}, {Chen}, {Chiang},
  {Chiang}, {Christensen}, {Church}, {Clements}, {Colombi}, {Colombo},
  {Couchot}, {Coulais}, {Crill}, {Curto}, {Cuttaia}, {Danese}, {Davies},
  {Davis}, {de Bernardis}, {de Rosa}, {de Zotti}, {Delabrouille}, {Delouis},
  {D{\'e}sert}, {Dickinson}, {Diego}, {Dolag}, {Dole}, {Donzelli}, {Dor{\'e}},
  {Douspis}, {Dunkley}, {Dupac}, {Efstathiou}, {Elsner}, {En{\ss}lin},
  {Eriksen}, {Finelli}, {Forni}, {Frailis}, {Fraisse}, {Franceschi}, {Gaier},
  {Galeotta}, {Galli}, {Ganga}, {Giard}, {Giardino}, {Giraud-H{\'e}raud},
  {Gjerl{\o}w}, {Gonz{\'a}lez-Nuevo}, {G{\'o}rski}, {Gratton}, {Gregorio},
  {Gruppuso}, {Gudmundsson}, {Haissinski}, {Hamann}, {Hansen}, {Hanson},
  {Harrison}, {Henrot-Versill{\'e}}, {Hern{\'a}ndez-Monteagudo}, {Herranz},
  {Hildebrand t}, {Hivon}, {Hobson}, {Holmes}, {Hornstrup}, {Hou}, {Hovest},
  {Huffenberger}, {Jaffe}, {Jaffe}, {Jewell}, {Jones}, {Juvela},
  {Keih{\"a}nen}, {Keskitalo}, {Kisner}, {Kneissl}, {Knoche}, {Knox}, {Kunz},
  {Kurki-Suonio}, {Lagache}, {L{\"a}hteenm{\"a}ki}, {Lamarre}, {Lasenby},
  {Lattanzi}, {Laureijs}, {Lawrence}, {Leach}, {Leahy}, {Leonardi},
  {Le{\'o}n-Tavares}, {Lesgourgues}, {Lewis}, {Liguori}, {Lilje},
  {Linden-V{\o}rnle}, {L{\'o}pez-Caniego}, {Lubin}, {Mac{\'\i}as-P{\'e}rez},
  {Maffei}, {Maino}, {Mand olesi}, {Maris}, {Marshall}, {Martin},
  {Mart{\'\i}nez-Gonz{\'a}lez}, {Masi}, {Massardi}, {Matarrese}, {Matthai},
  {Mazzotta}, {Meinhold}, {Melchiorri}, {Melin}, {Mendes}, {Menegoni},
  {Mennella}, {Migliaccio}, {Millea}, {Mitra}, {Miville-Desch{\^e}nes},
  {Moneti}, {Montier}, {Morgante}, {Mortlock}, {Moss}, {Munshi}, {Murphy},
  {Naselsky}, {Nati}, {Natoli}, {Netterfield}, {N{\o}rgaard-Nielsen},
  {Noviello}, {Novikov}, {Novikov}, {O'Dwyer}, {Osborne}, {Oxborrow}, {Paci},
  {Pagano}, {Pajot}, {Paladini}, {Paoletti}, {Partridge}, {Pasian},
  {Patanchon}, {Pearson}, {Pearson}, {Peiris}, {Perdereau}, {Perotto},
  {Perrotta}, {Pettorino}, {Piacentini}, {Piat}, {Pierpaoli}, {Pietrobon},
  {Plaszczynski}, {Platania}, {Pointecouteau}, {Polenta}, {Ponthieu}, {Popa},
  {Poutanen}, {Pratt}, {Pr{\'e}zeau}, {Prunet}, {Puget}, {Rachen}, {Reach},
  {Rebolo}, {Reinecke}, {Remazeilles}, {Renault}, {Ricciardi}, {Riller},
  {Ristorcelli}, {Rocha}, {Rosset}, {Roudier}, {Rowan-Robinson},
  {Rubi{\~n}o-Mart{\'\i}n}, {Rusholme}, {Sandri}, {Santos}, {Savelainen},
  {Savini}, {Scott}, {Seiffert}, {Shellard}, {Spencer}, {Starck}, {Stolyarov},
  {Stompor}, {Sudiwala}, {Sunyaev}, {Sureau}, {Sutton}, {Suur-Uski}, {Sygnet},
  {Tauber}, {Tavagnacco}, {Terenzi}, {Toffolatti}, {Tomasi}, {Tristram},
  {Tucci}, {Tuovinen}, {T{\"u}rler}, {Umana}, {Valenziano}, {Valiviita}, {Van
  Tent}, {Vielva}, {Villa}, {Vittorio}, {Wade}, {Wandelt}, {Wehus}, {White},
  {White}, {Wilkinson}, {Yvon}, {Zacchei}, \& {Zonca}}]{Planck13}
{Planck Collaboration} {et~al.}, 2014, \aap, 571, A16

\bibitem[{{Rauch} {et~al}\mbox{.}(2008){Rauch}, {Haehnelt}, {Bunker}, {Becker},
  {Marleau}, {Graham}, {Cristiani}, {Jarvis}, {Lacey}, {Morris}, {Peroux},
  {R{\"o}ttgering}, \& {Theuns}}]{Rauch08}
{Rauch} M. {et~al.}, 2008, \apj, 681, 856

\bibitem[{{Reuland} {et~al}\mbox{.}(2003){Reuland}, {van Breugel},
  {R{\"o}ttgering}, {de Vries}, {Stanford}, {Dey}, {Lacy}, {Bland-Hawthorn},
  {Dopita}, \& {Miley}}]{Reuland03}
{Reuland} M. {et~al.}, 2003, \apj, 592, 755

\bibitem[{{Rosdahl} \& {Blaizot}(2012)}]{Rosdahl12}
{Rosdahl} J., {Blaizot} J., 2012, \mnras, 423, 344

\bibitem[{{Rosdahl} {et~al}\mbox{.}(2013){Rosdahl}, {Blaizot}, {Aubert},
  {Stranex}, \& {Teyssier}}]{Rosdahl13}
{Rosdahl} J., {Blaizot} J., {Aubert} D., {Stranex} T., {Teyssier} R., 2013,
  \mnras, 436, 2188

\bibitem[{{Rosdahl} {et~al}\mbox{.}(2018){Rosdahl}, {Katz}, {Blaizot}, {Kimm},
  {Michel-Dansac}, {Garel}, {Haehnelt}, {Ocvirk}, \& {Teyssier}}]{Rosdahl18}
{Rosdahl} J. {et~al.}, 2018, \mnras, 479, 994

\bibitem[{{Rosdahl} \& {Teyssier}(2015)}]{Rosdahl15}
{Rosdahl} J., {Teyssier} R., 2015, \mnras, 449, 4380

\bibitem[{{Rosen} \& {Bregman}(1995)}]{rosen95}
{Rosen} A., {Bregman} J.~N., 1995, \apj, 440, 634

\bibitem[{{Scannapieco}(2017)}]{Scannapieco17}
{Scannapieco} E., 2017, \apj, 837, 28

\bibitem[{{Shapley} {et~al}\mbox{.}(2003){Shapley}, {Steidel}, {Pettini}, \&
  {Adelberger}}]{Shapley03}
{Shapley} A.~E., {Steidel} C.~C., {Pettini} M., {Adelberger} K.~L., 2003, \apj,
  588, 65

\bibitem[{{Smit} {et~al}\mbox{.}(2017){Smit}, {Swinbank}, {Massey}, {Richard},
  {Smail}, \& {Kneib}}]{Smit17}
{Smit} R., {Swinbank} A.~M., {Massey} R., {Richard} J., {Smail} I., {Kneib}
  J.~P., 2017, \mnras, 467, 3306

\bibitem[{{Smith} {et~al}\mbox{.}(2019){Smith}, {Ma}, {Bromm}, {Finkelstein},
  {Hopkins}, {Faucher-Gigu{\`e}re}, \& {Kere{\v{s}}}}]{Smith19}
{Smith} A., {Ma} X., {Bromm} V., {Finkelstein} S.~L., {Hopkins} P.~F.,
  {Faucher-Gigu{\`e}re} C.-A., {Kere{\v{s}}} D., 2019, \mnras, 484, 39

\bibitem[{{Smith} {et~al}\mbox{.}(2015){Smith}, {Safranek-Shrader}, {Bromm}, \&
  {Milosavljevi{\'c}}}]{Smith15}
{Smith} A., {Safranek-Shrader} C., {Bromm} V., {Milosavljevi{\'c}} M., 2015,
  \mnras, 449, 4336

\bibitem[{{Song} {et~al}\mbox{.}(2020){Song}, {Seon}, \& {Hwang}}]{Song20}
{Song} H., {Seon} K.-I., {Hwang} H.~S., 2020, \apj, 901, 41

\bibitem[{{Stanway} {et~al}\mbox{.}(2016){Stanway}, {Eldridge}, \&
  {Becker}}]{Stanway16}
{Stanway} E.~R., {Eldridge} J.~J., {Becker} G.~D., 2016, \mnras, 456, 485

\bibitem[{{Steidel} {et~al}\mbox{.}(2011){Steidel}, {Bogosavljevi{\'c}},
  {Shapley}, {Kollmeier}, {Reddy}, {Erb}, \& {Pettini}}]{Steidel11}
{Steidel} C.~C., {Bogosavljevi{\'c}} M., {Shapley} A.~E., {Kollmeier} J.~A.,
  {Reddy} N.~A., {Erb} D.~K., {Pettini} M., 2011, \apj, 736, 160

\bibitem[{{Steidel} {et~al}\mbox{.}(2010){Steidel}, {Erb}, {Shapley},
  {Pettini}, {Reddy}, {Bogosavljevi{\'c}}, {Rudie}, \& {Rakic}}]{Steidel10}
{Steidel} C.~C., {Erb} D.~K., {Shapley} A.~E., {Pettini} M., {Reddy} N.,
  {Bogosavljevi{\'c}} M., {Rudie} G.~C., {Rakic} O., 2010, \apj, 717, 289

\bibitem[{{Suresh} {et~al}\mbox{.}(2019){Suresh}, {Nelson}, {Genel}, {Rubin},
  \& {Hernquist}}]{Suresh19}
{Suresh} J., {Nelson} D., {Genel} S., {Rubin} K. H.~R., {Hernquist} L., 2019,
  \mnras, 483, 4040

\bibitem[{{Swinbank} {et~al}\mbox{.}(2007){Swinbank}, {Bower}, {Smith},
  {Wilman}, {Smail}, {Ellis}, {Morris}, \& {Kneib}}]{Swinbank07}
{Swinbank} A.~M., {Bower} R.~G., {Smith} G.~P., {Wilman} R.~J., {Smail} I.,
  {Ellis} R.~S., {Morris} S.~L., {Kneib} J.~P., 2007, \mnras, 376, 479

\bibitem[{{Tapken} {et~al}\mbox{.}(2007){Tapken}, {Appenzeller}, {Noll},
  {Richling}, {Heidt}, {Meink{\"o}hn}, \& {Mehlert}}]{Tapken07}
{Tapken} C., {Appenzeller} I., {Noll} S., {Richling} S., {Heidt} J.,
  {Meink{\"o}hn} E., {Mehlert} D., 2007, \aap, 467, 63

\bibitem[{{Tasitsiomi}(2006)}]{Tasitsiomi06}
{Tasitsiomi} A., 2006, \apj, 645, 792

\bibitem[{{Teyssier}(2002)}]{Teyssier02}
{Teyssier} R., 2002, \aap, 385, 337

\bibitem[{{Thornton} {et~al}\mbox{.}(1998){Thornton}, {Gaudlitz}, {Janka}, \&
  {Steinmetz}}]{Thornton98}
{Thornton} K., {Gaudlitz} M., {Janka} H.~T., {Steinmetz} M., 1998, \apj, 500,
  95

\bibitem[{{Trainor} {et~al}\mbox{.}(2015){Trainor}, {Steidel}, {Strom}, \&
  {Rudie}}]{Trainor15}
{Trainor} R.~F., {Steidel} C.~C., {Strom} A.~L., {Rudie} G.~C., 2015, \apj,
  809, 89

\bibitem[{{Trebitsch} {et~al}\mbox{.}(2017){Trebitsch}, {Blaizot}, {Rosdahl},
  {Devriendt}, \& {Slyz}}]{Trebitsch17}
{Trebitsch} M., {Blaizot} J., {Rosdahl} J., {Devriendt} J., {Slyz} A., 2017,
  \mnras, 470, 224

\bibitem[{{Tweed} {et~al}\mbox{.}(2009){Tweed}, {Devriendt}, {Blaizot},
  {Colombi}, \& {Slyz}}]{Tweed09}
{Tweed} D., {Devriendt} J., {Blaizot} J., {Colombi} S., {Slyz} A., 2009, \aap,
  506, 647

\bibitem[{{van de Voort} {et~al}\mbox{.}(2019){van de Voort}, {Springel},
  {Mandelker}, {van den Bosch}, \& {Pakmor}}]{vandevoort19}
{van de Voort} F., {Springel} V., {Mandelker} N., {van den Bosch} F.~C.,
  {Pakmor} R., 2019, \mnras, 482, L85

\bibitem[{{Verhamme} {et~al}\mbox{.}(2012){Verhamme}, {Dubois}, {Blaizot},
  {Garel}, {Bacon}, {Devriendt}, {Guiderdoni}, \& {Slyz}}]{Verhamme12}
{Verhamme} A., {Dubois} Y., {Blaizot} J., {Garel} T., {Bacon} R., {Devriendt}
  J., {Guiderdoni} B., {Slyz} A., 2012, \aap, 546, A111

\bibitem[{{Verhamme} {et~al}\mbox{.}(2018){Verhamme}, {Garel}, {Ventou},
  {Contini}, {Bouch{\'e}}, {Herenz}, {Richard}, {Bacon}, {Schmidt}, {Maseda},
  {Marino}, {Brinchmann}, {Cantalupo}, {Caruana}, {Cl{\'e}ment}, {Diener},
  {Drake}, {Hashimoto}, {Inami}, {Kerutt}, {Kollatschny}, {Leclercq},
  {Patr{\'\i}cio}, {Schaye}, {Wisotzki}, \& {Zabl}}]{Verhamme18}
{Verhamme} A. {et~al.}, 2018, \mnras, 478, L60

\bibitem[{{Verhamme} {et~al}\mbox{.}(2008){Verhamme}, {Schaerer}, {Atek}, \&
  {Tapken}}]{Verhamme08}
{Verhamme} A., {Schaerer} D., {Atek} H., {Tapken} C., 2008, \aap, 491, 89

\bibitem[{{Verhamme} {et~al}\mbox{.}(2006){Verhamme}, {Schaerer}, \&
  {Maselli}}]{Verhamme06}
{Verhamme} A., {Schaerer} D., {Maselli} A., 2006, \aap, 460, 397

\bibitem[{{Wang}(1995)}]{Wang95}
{Wang} B., 1995, \apj, 444, 590

\bibitem[{{Wisotzki} {et~al}\mbox{.}(2016){Wisotzki}, {Bacon}, {Blaizot},
  {Brinchmann}, {Herenz}, {Schaye}, {Bouch{\'e}}, {Cantalupo}, {Contini},
  {Carollo}, {Caruana}, {Courbot}, {Emsellem}, {Kamann}, {Kerutt}, {Leclercq},
  {Lilly}, {Patr{\'\i}cio}, {Sandin}, {Steinmetz}, {Straka}, {Urrutia},
  {Verhamme}, {Weilbacher}, \& {Wendt}}]{Wisotzki16}
{Wisotzki} L. {et~al.}, 2016, \aap, 587, A98

\bibitem[{{Wisotzki} {et~al}\mbox{.}(2018){Wisotzki}, {Bacon}, {Brinchmann},
  {Cantalupo}, {Richter}, {Schaye}, {Schmidt}, {Urrutia}, {Weilbacher},
  {Akhlaghi}, {Bouch{\'e}}, {Contini}, {Guiderdoni}, {Herenz}, {Inami},
  {Kerutt}, {Leclercq}, {Marino}, {Maseda}, {Monreal-Ibero}, {Nanayakkara},
  {Richard}, {Saust}, {Steinmetz}, \& {Wendt}}]{Wisotzki18}
{Wisotzki} L. {et~al.}, 2018, \nat, 562, 229

\bibitem[{{Wofford} {et~al}\mbox{.}(2013){Wofford}, {Leitherer}, \&
  {Salzer}}]{Wofford13}
{Wofford} A., {Leitherer} C., {Salzer} J., 2013, \apj, 765, 118

\bibitem[{{Yusef-Zadeh} {et~al}\mbox{.}(1984){Yusef-Zadeh}, {Morris}, \&
  {White}}]{YusefZadeh84}
{Yusef-Zadeh} F., {Morris} M., {White} R.~L., 1984, \apj, 278, 186

\bibitem[{{Zheng} {et~al}\mbox{.}(2010){Zheng}, {Cen}, {Trac}, \&
  {Miralda-Escud{\'e}}}]{Zheng10}
{Zheng} Z., {Cen} R., {Trac} H., {Miralda-Escud{\'e}} J., 2010, \apj, 716, 574

\bibitem[{{Zheng} {et~al}\mbox{.}(2011){Zheng}, {Cen}, {Weinberg}, {Trac}, \&
  {Miralda-Escud{\'e}}}]{Zheng11}
{Zheng} Z., {Cen} R., {Weinberg} D., {Trac} H., {Miralda-Escud{\'e}} J., 2011,
  \apj, 739, 62

\bibitem[{{Zheng} \& {Miralda-Escud{\'e}}(2002)}]{Zheng02}
{Zheng} Z., {Miralda-Escud{\'e}} J., 2002, \apj, 578, 33

\end{thebibliography}
